\documentclass[usenatbib,usegraphicx]{mn2e}

\usepackage{pslatex,color,tabularx,float,multicol,caption}
\usepackage{amsmath,amsfonts,amssymb,amsxtra,eufrak,url}

\def\idm#1{{\mbox{\scriptsize #1}}}

\def\vec#1{{\pmb #1}}

\newcommand{\au}{\mbox{au}}
\newcommand{\mS}{\mbox{M}_{\odot}}
\newcommand{\mE}{\mbox{M}_{\oplus}}

\newcommand{\RE}{\mbox{R}_{\oplus}}
\newcommand{\Mmean}{\mathcal{M}}

\newcommand{\Deltaaph}{\Delta_{\idm{a-ph}}}
\newcommand{\TOC}{T_{\idm{(O-C)}}}
\newcommand{\Chi}{\chi^2_{\nu}}
\def\kepler{{\em Kepler}}

\title[Periodic configuration of Kepler-25?]
{A periodic configuration of the Kepler-25 planetary system?}
\author[Cezary Migaszewski \& Krzysztof Go\'zdziewski]{Cezary Migaszewski$^{1\star}$ \& Krzysztof Go\'zdziewski$^{1\star}$\\
$^{1}$Centre for Astronomy, Faculty of Physics, Astronomy and Informatics, Nicolaus Copernicus University, Grudziadzka 5, 87-100 Toru\'n, Poland}
\begin{document}
%
\date{Accepted .... Received ...; in original form ...}
\pagerange{\pageref{firstpage}--\pageref{lastpage}} \pubyear{2018}
\maketitle
\label{firstpage}

\captionsetup[figure]{labelfont=bf,font=small}

\begin{abstract}

We study proximity of the Kepler-25 planetary system to a periodic configuration, which is known to be the final state of a system that undergoes smooth migration resulting from the planet-disc interaction. We show that the system is close to the periodic configuration of 2:1 mean motion resonance (MMR) what indicates that its past migration was neither disturbed significantly by turbulence in the disc nor the orbits were perturbed by planetesimals that left after the disc dispersal. We show that, because of the TTV model degeneracy, a periodic configuration is difficult to be found when the standard modelling of the transit timing variations (TTVs) is used. The TTV signal of a periodic configuration (with anti-aligned apsidal lines) may be misinterpreted as an aligned non-resonant system. We demonstrate that the standard MCMC modelling of the Kepler-25 TTVs is very sensitive to an a~priori information on the eccentricities (prior probability distributions). Wide priors (of the order of the ones typically used in the literature) result in favouring the aligned non-resonant configurations with small planets' masses and moderate eccentricities, while for the narrower priors the most likely are the anti-aligned resonant systems with larger masses and very low eccentricities. 

\end{abstract}


\section{Introduction}

{\let\thefootnote\relax\footnotetext{$^{\star}$Email: migaszewski@umk.pl (CM), chris@umk.pl (KG)}}

{It is known that a periodic configuration (in a reference frame co-rotating with the apsidal lines) is one of the possible outcomes of the smooth disc-induced migration of a two-planet system \citep{Hadjidemetriou2006,Migaszewski2015}.} The migration can be alternately convergent or divergent during the whole lifetime of the disc. Depending on particular history of the migration, the final period ratio $P_2/P_1$ (where $P_1, P_2$ are the orbital periods of the inner and the outer planet, respectively) may be close to a nominal value of a particular mean motion resonance (e.g., 2/1, 3/2, 4/3, etc.) or shifted away from such a value. The proximity of the final system to the periodic configuration is, however, not related to the final period ratio. The system may have $P_2/P_1$ very distant from the resonant value but still be a periodic configuration, i.e., be resonant in terms of the resonant angles librations.

{Nevertheless, as mentioned above, the periodic configuration is not the only possible result of the migration. The system can deviate from periodicity if the migration is too rapid (i.e., the evolution is non-adiabatic) or if the system passes through the resonance. The latter may happen if the migration is divergent or if the resonance capture is only temporary, i.e., librations around the periodic configuration (an equilibrium in the averaged model of the resonance) are overstable \citep{Goldreich2014}. It is also possible that the amplitude of the librations saturates at a non-zero value, resulting in a resonant system that is shifted away from the periodic configuration \citep{Goldreich2014,Deck2015}. Although, in general, the migration does not necessarily lead to the periodic configuration, a system that is close-to-periodic has been likely formed on the way of migration.}

This work is related to the problem of explaining the observed period ratio distribution of the KEPLER systems \citep{Fabrycky2014}. Only small fraction of multi-planet systems have $P_2/P_1$ close to resonant values. There are several explanations of how a given resonant pair moved away from the resonance. They may be divided into two groups. In the scenarios from the first group the system is being moved away from the resonance because of perturbations resulting from remnant planetesimals \citep{Chatterjee2015,Rein2012} or the interaction with a turbulent disc \citep{Nelson2005}. In the scenarios from the second group the migration is smooth (i.e., not disturbed by the forces mentioned above) but not necessarily convergent during the whole evolution of the system. The divergence of the migration may result from particular physical conditions in an evolving disc \citep{Migaszewski2015} or from the tidal star-planet interaction \citep{Papaloizou2010,Papaloizou2011,Batygin2013,Delisle2014a,Delisle2014b}. The latter mechanism acts, however, only for very short-period planets ($P \sim 1\,$day). Another mechanism, that can be counted to this group has been recently proposed \citep{Ramos2017}. They show that the equilibrium values of the period ratio of two short-period planets may differ significantly from the nominal resonant values if a disc in which they migrated has a small aspect ratio and is significantly flared.
\cite{Charalambous2017} studied the Kepler-25 system with two transiting planets \citep[discovered by the \kepler{} mission;][]{Steffen2012} in this context, and showed that its period ratio of $2.039$ may result from the migration in the disc of properties suggested by \cite{Ramos2017}.

The first group scenarios result in not only the $P_2/P_1$ deviation from the resonant value but also in the deviation of the system from the periodic configuration. As discussed in \citep{Rein2009} stochastic forces acting on a planet affects both the semi-major axis $a$ and the eccentricity $e$ in similar magnitude. That means that if the Kepler-25 with $P_2/P_1 \approx 2.039$ was shifted away from the 2:1 MMR by such forces, the variation of the eccentricities should be $\gtrsim 0.01$, which is far from the periodic system, as $e_1 \sim 10^{-3}$ and $e_2 \sim 10^{-4}$ for the periodic configuration of $P_2/P_1 = 2.039$ and the planets' masses in super-Earths regime \citep[as the measured planets' radii, $R_1 = (2.64 \pm 0.04)\,\RE$ and $R_2 = (4.51 \pm 0.08)\,\RE$ suggest;][]{Rowe2015}. [We use the indices of $1$ and $2$ for the inner and the outer planet, instead of b and~c.] That means that the Kepler-25 system is a good tester of the migration as a formation mechanism of configurations with $P_2/P_1$ relatively close to (but not exactly at) the resonant values.

This work is organized as follows. In Section~2 we study branches of periodic configurations as a function of the planets' masses $m_1, m_2$ (for the inner and the outer planets, respectively) and the period ratio. As the eccentricities $e_1, e_2$ depend on $m_1, m_2, P_2/P_1$ and the TTV amplitudes depend on $e_1, e_2$, we show that it is possible to constrain the masses of the Kepler-25 system, when assuming a periodic configuration of this system. The periodic configuration fitting procedure is presented in Section~3. In the next section we try to verify whether or not a system which is far from a periodic configuration may be misinterpreted as a periodic system, when applying the procedure explained in Section~3. On the other hand, in Section~5 we try to find out if a periodic configuration may be misinterpreted as far form periodic when applying the standard TTV fitting procedure. The last section is devoted to summary and conclusions.

\section{Migration and periodic configurations}

Families of periodic orbits of two-planet systems has been widely studied for different MMRs and in a wide range of planets' masses \citep[e.g.,][]{Hadjidemetriou1975,Hadjidemetriou2006,Antoniadou2014}. Their connection with the migration has been also pointed out \citep[e.g.,][]{Beauge2006,Hadjidemetriou2010,Migaszewski2015}. Father in this work we show the connection between the periodic configurations and the TTVs.

When using the averaging approach to the resonant two-planet system \citep[e.g.,][]{Beauge2003b}, a periodic configuration corresponds to an equilibrium in a reference frame co-rotating with the apsidal lines. For a stable equilibrium (which is the case we are interested in) the eccentricities, semi-major axes, the difference of the longitudes of pericentres ($\Delta\varpi = \varpi_1 - \varpi_2$, where $\varpi_i$ is the longitude of the $i$-th planet's pericentre) and the resonant angles are constant. The individual $\varpi_i$, on contrary, vary linearly in time. 
{The so-called free eccentricities of a system in an equilibrium equal zero and the apsidal lines rotate with a period that equals the so-called super-period \citep{Lithwick2012}, $T_q = |q/P_1 - (q+1)/P_2|^{-1}$, for the $(q+1)$:$q$ resonance.}

From the observational point of view, a uniformly rotating orbit of given (and fixed) $a$ and $e$ should lead to a periodic signal in TTV, as an actual orientation of the orbit determines whether the transit occurs earlier or later than it stems from the Keplerian motion of the planet. As the true anomaly $\nu$ may be expressed for $e \ll 1$ by the mean anomaly $\Mmean$ as $\nu \approx \Mmean + 2 e \sin\Mmean$ \citep[e.g.,][]{Brouwer1961}, the TTV should be sinusoidal (for low $e$) with a semi-amplitude $A$ that depends on $e$ through the relation $A/P = (2 e)/(2\pi)$. If the orbits of a periodic two-planet configuration are anti-aligned, $\Delta\varpi = \pi$, the TTV signals for the planets should be in anti-phase. If the orbits are aligned, $\Delta\varpi = 0$, the signals are in phase, while for $\Delta\varpi$ different from $0$ or $\pi$ \citep[the asymmetric co-rotation, e.g.,][]{Beauge2003} the difference in phases of the signals are neither $0$ nor $\pi$.

\begin{figure}
\centerline{
\vbox{
\hbox{\includegraphics[width=0.42\textwidth]{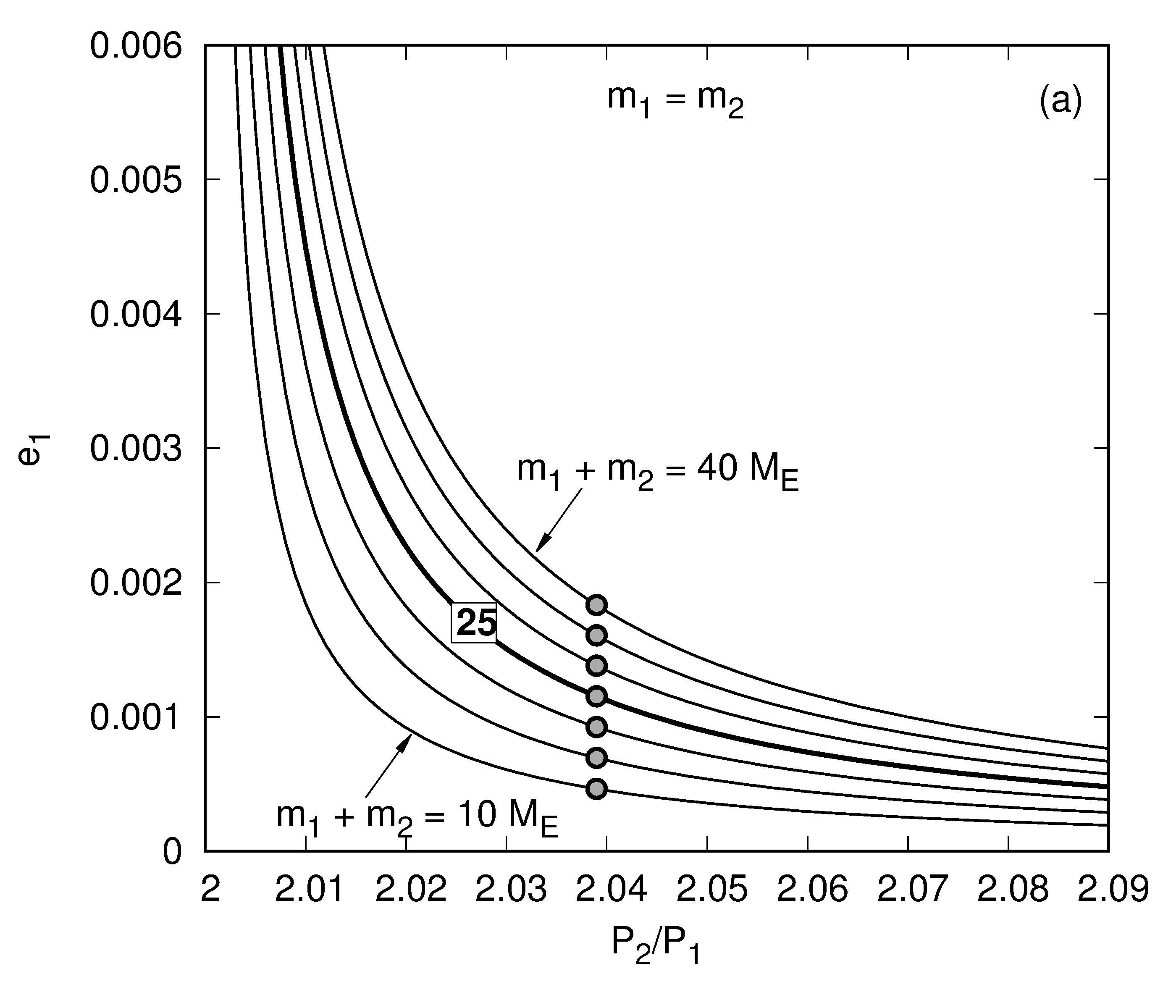}}
\hbox{\includegraphics[width=0.42\textwidth]{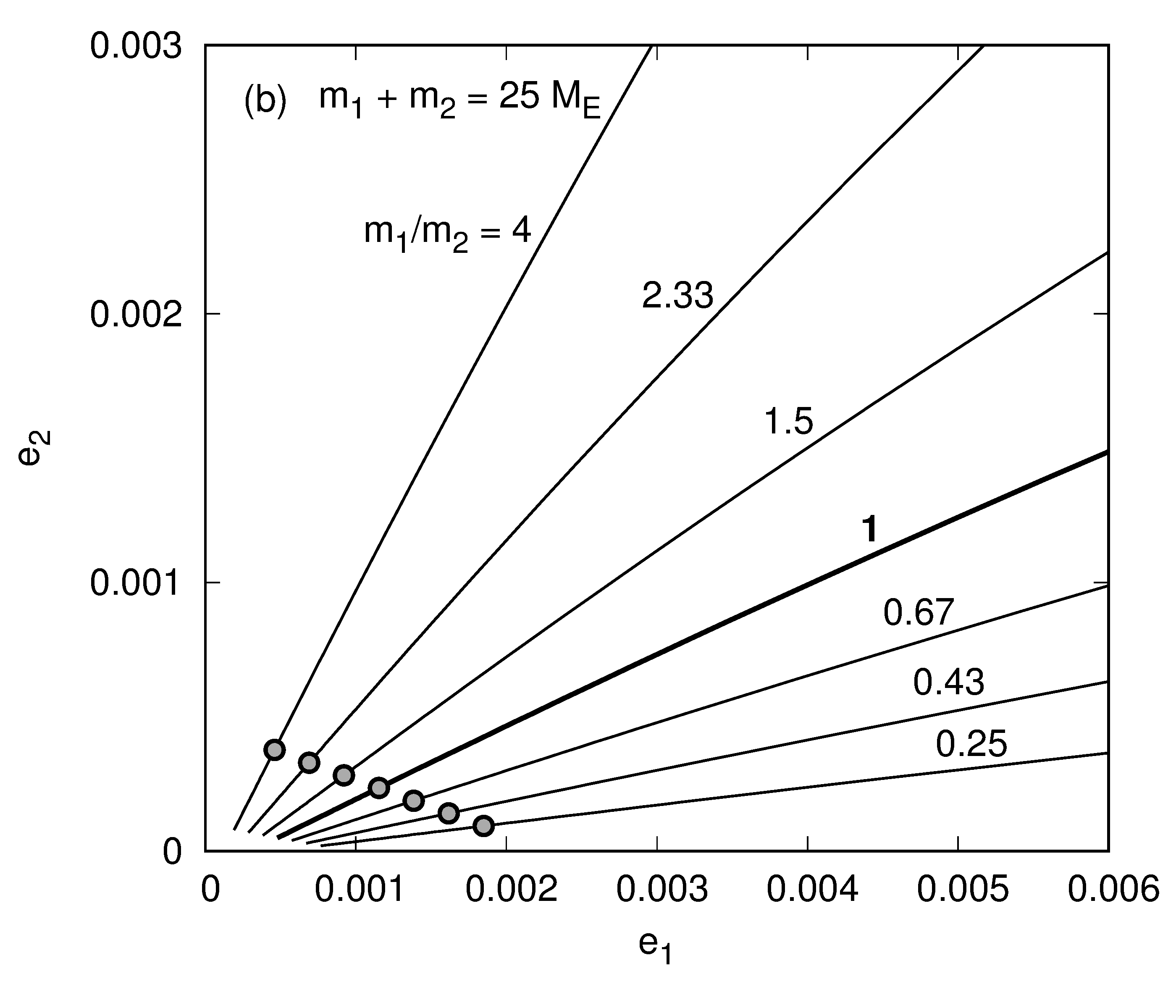}}
\hbox{\includegraphics[width=0.42\textwidth]{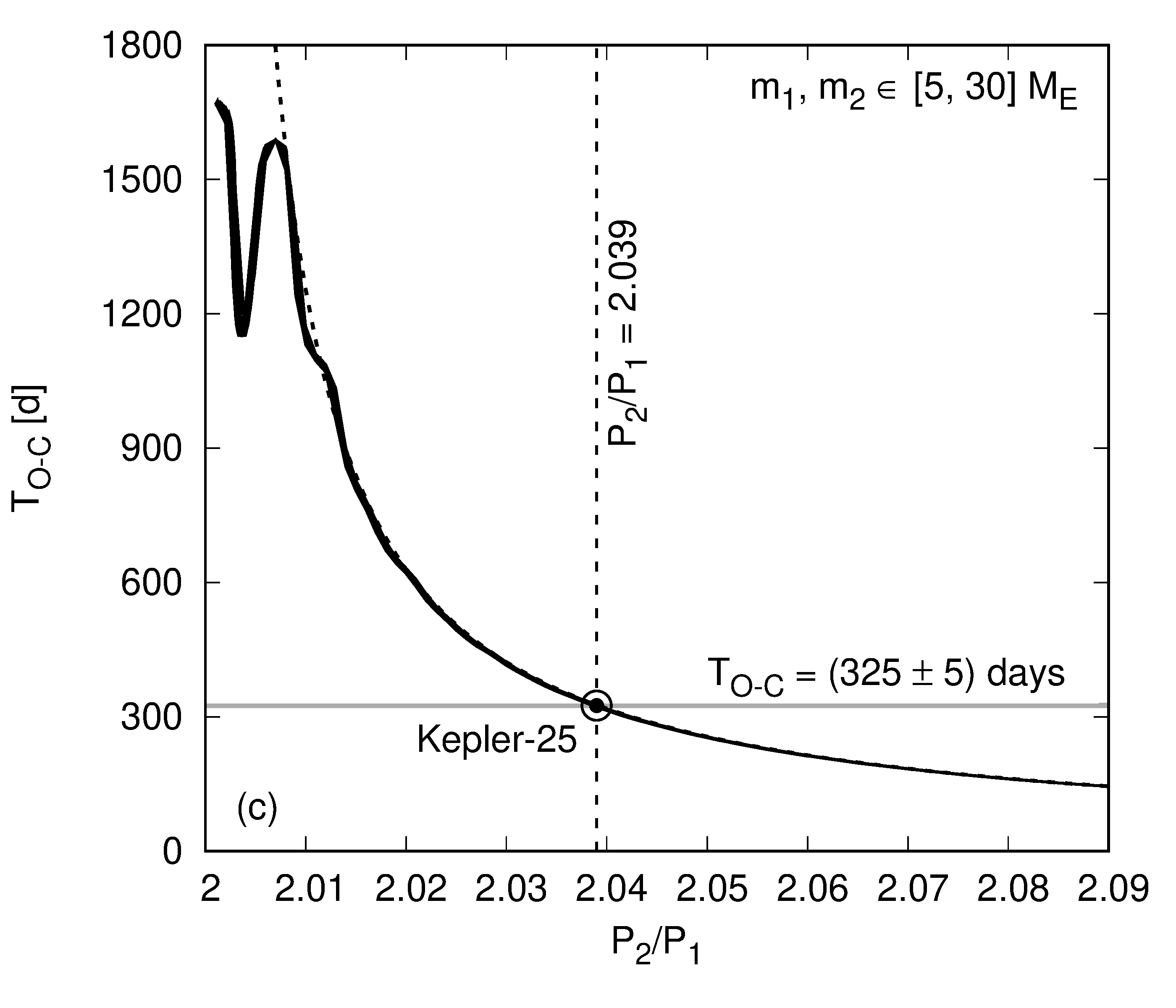}}
}
}
\caption{Branches of periodic configurations for different planets' masses {(the mass of the star is fixed at the value corresponding to Kepler-25, i.e., $m_0=1.19\,\mS$)}. Panel~(a) presents the results for fixed mass ratio (equal $1$) and different $m_1 + m_2$ (the values vary from $10\,\mE$ to $40\,\mE$, with a $5\,\mE$ increment) at a plane of $(P_2/P_1, e_1)$. Panel~(b) shows the results at the eccentricities plane obtained for a fixed sum of the masses and the ratio $m_1/m_2$ being varied in a range of $(0.25, 4)$. Big gray dots indicate $(P_2/P_1)=2.039$ that corresponds to the Kepler-25 system. Panel~(c) illustrates the period of the TTV signal, $\TOC$, as a function of the orbital period ratio computed for periodic configurations of both the planets' masses in a range of $[5, 30]\,\mE$. {An analytic estimate of $\TOC$, the super-period, is shown with a dashed curve.} The circle symbol indicate the position of the Kepler-25 system at the plot.}
\label{fig:periodic}
\end{figure}

As the amplitudes of TTV depend on the eccentricities, a natural question arises on how the eccentricities of periodic configurations depend on planets' masses. Figure~\ref{fig:periodic} presents branches (families) of periodic orbits obtained for different $m_1, m_2$ in certain range of $P_2/P_1 > 2$. The procedure of finding periodic orbits is the following \citep[e.g.,][]{Migaszewski2017}. For given masses we search for such $\vec{p} = (a_1\cos\Mmean_1, a_1\sin\Mmean_1, a_2\cos\Mmean_2, a_2\sin\Mmean_2, e_1\cos\Delta\varpi, e_1\sin\Delta\varpi, e_2)$ that satisfies {$\delta = 0$, where $\delta \equiv \| \vec{p}(t = T) - \vec{p}(t = 0)\|$, and $T$ is the period}.

For 2:1~MMR the period $T$ corresponds to one revolution of the outer planet and two revolutions of the inner planet. The stable equilibrium (when we use the averaged model of the resonance) of 2:1~MMR in a regime of small eccentricities exist for $\Delta\varpi = \pi$ and values of the resonant angles depend on $P_2/P_1$. For $P_2/P_1 > 2$ the angles defined as $\phi_1 \equiv \lambda_1 - 2 \lambda_2 + \varpi_1$ and $\phi_2 \equiv \lambda_1 - 2 \lambda_2 + \varpi_2$ (where $\lambda_i \equiv \Mmean_i + \varpi_i$ is the mean longitude of $i$-th planet) equal $0$ and $\pi$, respectively, while for $P_2/P_1 < 2$, $\phi_1 = \pi$ and $\phi_2 = 0$. The Kepler-25 system has $P_2/P_1 = 2.039$, thus we will consider only the first case. Naturally, saying that the equilibrium of the averaged system corresponds to $\Delta\varpi = \pi$ and $\phi_1 = 0$ means that for the unaveraged system $\Delta\varpi$ and $\phi_1$ oscillate around those values. The amplitudes of the oscillations depend on the distance of $P_2/P_1$ to the resonant values of particular resonance \citep{Migaszewski2015}. The closer $P_2/P_1$ is to the nominal value of given MMR, the smaller are the amplitudes.

When searching for a periodic configuration one needs to set initial values of the mean anomalies, such that $\Delta\varpi = \pi$ and $\phi_1 = 0$. There are two combinations of $(\Mmean_1, \Mmean_2)$ that satisfy the condition, i.e., $(0, 0)$ and $(0, \pi)$. The results presented in Fig.~\ref{fig:periodic} were obtained for the former choice of the mean anomalies. When the angles $\Mmean_1, \Mmean_2, \Delta\varpi$ are set, we search for $\vec{p}(t=0)$ that gives $\delta=0$ in the parameter space of $(P_2/P_1, e_1, e_2)$. As it is known \citep[e.g.,][]{Hadjidemetriou2006} there exists a curve satisfying $\delta=0$. In the regime of $(P_2/P_1, e_1, e_2)$, that we are interested in, the curve can be parametrized with $P_2/P_1$, i.e., $e_1 = e_1 (P_2/P_1)$ and $e_2 = e_2 (P_2/P_1)$. Therefore, in order to find a family of periodic configurations one needs to search for $\delta(e_1, e_2) = 0$ for a series of values of $P_2/P_1$ in a given range. The equation $\delta(e_1, e_2) = 0$ is being solved numerically with a help of the Powell's method combined with the golden section \citep[e.g.,][]{Press2002}.

Figure~\ref{fig:periodic}a presents the families of periodic orbits obtained for $P_2/P_1 \in (2, 2.09]$ and for various $m_1, m_2$ with their ratio kept constant, $m_1/m_2 = 1$. Each curve, presented in the $(P_2/P_1, e_1)$-plane corresponds to different value of $m_1 + m_2$. When the masses are lower, $e_1$ (and similarly $e_2$) are lower for given $P_2/P_1$. The period ratio value of Kepler-25, $P_2/P_1 = 2.039$, is marked with large gray dots. 
Panel~(b) of Fig.~\ref{fig:periodic} presents the branches of periodic configurations in the eccentricities plane. Here, the sum of the masses is fixed, while the ratio varies. Clearly, when $m_1/m_2$ increases (for a given $P_2/P_1$) $e_1$ decreases, while $e_2$ increases.

As $A_i \propto e_i$, by repeating the analysis for different $m_1, m_2$ one can find an approximate relation between the amplitudes $A_1, A_2$ and the masses, i.e., $A_i = A_i (m_1, m_2; P_2/P_1)$. Roughly speaking, $A_1 \propto m_2$ and $A_2 \propto m_1$ in a regime of small masses. Therefore, it should be possible, in principle, to constrain the planets' masses by fitting the model of a periodic configuration to the TTV data (naturally, if the real system is close-to-periodic, what we demonstrate later in this work). The fitting procedure is presented in the next section.

In contrast with the amplitudes dependence on the planets' masses, a period of the rotation of the system as a whole (that equals to the (O-C)-signal period, $\TOC$) does not depend on the masses significantly (in a small mass regime). Figure~\ref{fig:periodic}c presents $\TOC$ as a function of $P_2/P_1$ for periodic configurations found for the masses in a range of $[5,30]\,\mE$. We note that the $\TOC$ dependence on the period ratio agrees with the analytic model of \cite{Lithwick2012}, {i.e., the super-period, that equals $|1/P_1 - 2/P_2|^{-1}$ for the 2:1~MMR (the analytic prediction is marked with a dashed curve in Fig.~\ref{fig:periodic}c).} The Kepler-25 system, whose position at the plot is marked with a circle symbol, lies exactly at the curve corresponding to the families of periodic configurations. {Nevertheless, the agreement between $\TOC$ of Kepler-25 and the super-period does not mean that the system has to be a periodic configuration, as the super-period corresponds to the variation of the longitude of conjunction \citep{Lithwick2012}.}

\section{Modelling TTV with periodic configurations}

We use the TTV data from the catalogue of \cite{Rowe2015} and search for the best-fitting parameters of the two-planet model in terms of the minimum of the standard $\Chi$ function. What differs the procedure from the standard fitting approach is that here the orbital parameters $(a_1, a_2, e_1, e_2, \varpi_1, \varpi_2, \Mmean_1, \Mmean_2)$ given at time $t_0$ are not free parameters. Each configuration for which $\Chi$ is being evaluated has to be periodic. Therefore, the optimization occurs at a hyper-surface embedded in the parameters space.

As we already showed in the previous section, the periodic configurations form a family parametrized by the period ratio [for given initial values of $(\Delta\varpi, \Mmean_1, \Mmean_2)$; we will use $(\pi, 0, 0)$] and the planets' masses. Particular values of the eccentricities as well as the angles ($\Delta\varpi, \Mmean_1, \Mmean_2$) are functions of time, i.e., they depend on the phase of the periodic evolution, namely $\tau \in [0, T)$. Although the dynamics of the system in scalable in a sense of physical dimensions as well as it is rotation invariant, when modelling the observations one needs to find an appropriate scale (that is given by initial $P_1$) and orientation. For the TTV analysis the only Euler angle that needs to be fitted is the angle that measures the rotation in the orbital plane (as we assume that the inclination $I=\pi/2$ and the TTV signal is invariant with respect to the rotation in the sky plane; formally we put $\Omega=0$ for both longitudes of the ascending nodes). We denote the Euler angle by $\varpi_0$, and for a given configuration tested in the fitting procedure this value is being added to both longitudes of pericentre. 

Finally, the complete set of free parameters of the model $\vec{x} = (m_1, m_2, P_2/P_1, P_1, \tau, \varpi_0)$. Direct minimizing $\Chi = \Chi(\vec{x})$ would be too long, as finding a periodic configuration for a given $(m_1, m_2, P_2/P_1)$ occurs in a numerical process of solving $\delta(e_1, e_2) = 0$. In order to make the procedure work efficiently, we treat the first two parameters $(m_1, m_2)$ as fixed in a given fitting process, while the fitting is being repeated for subsequent points at the $(m_1, m_2)$-plane, taken from a grid, namely $m_i \in [5, 20]\,\mE$, with an increment of $0.75\,\mE$.

The period ratio (that has an osculating sense) is being chosen before a given TTV fitting process in such a way that the ratio of the periods of the transit times series $\{n, t_n\}$ agrees with the observational value. Speaking in more details, for a given initial set of osculating Keplerian elements, the equations of motion are solved numerically, the transit times for both the planets are found and the linear model of TTs, i.e., $T_n = T_0 + n \langle P \rangle$ is fitted to the $\{n, t_n\}$ series. Next, the value of $\langle P_2 \rangle / \langle P_1 \rangle$ is compared to the observational value of the Kepler-25 system, i.e., $\approx 2.039$. The scale of the system is being found in the same way, i.e., $\langle P_1 \rangle$ needs to equal the observational value of $\approx 6.2385\,$d.

\begin{figure}
\centerline{
\vbox{
\hbox{
\includegraphics[width=0.48\textwidth]{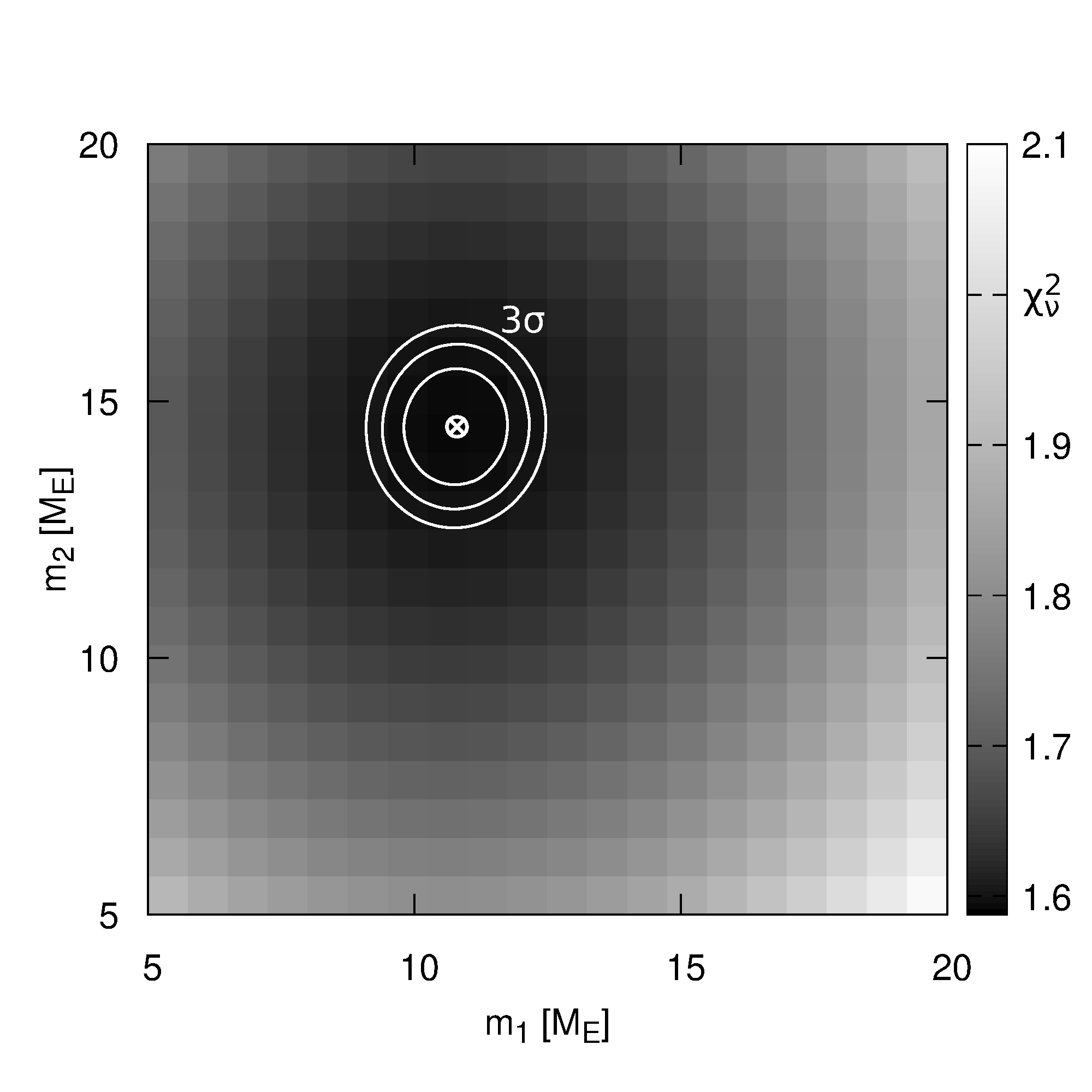}
}
}
}
\caption{A $\Chi$-scan at the $(m_1, m_2)$-plane. See the text for details. Contours indicate $1-, 2-$~and $3-\sigma$ confidence levels.}
\label{fig:best_fit_scan}
\end{figure}

Therefore, for given $(m_1, m_2)$ the $\Chi$ is being minimized in the two-paremeter space, i.e., $\Chi = \Chi(\tau, \varpi_0)$. We use the same numerical optimization scheme as in searching for the periodic orbits, described in the previous section.  Figure~\ref{fig:best_fit_scan} presents a scan of $\Chi$ computed at a grid of masses, as described above. 

\begin{figure}
\centerline{
\vbox{
\hbox{
\includegraphics[width=0.48\textwidth]{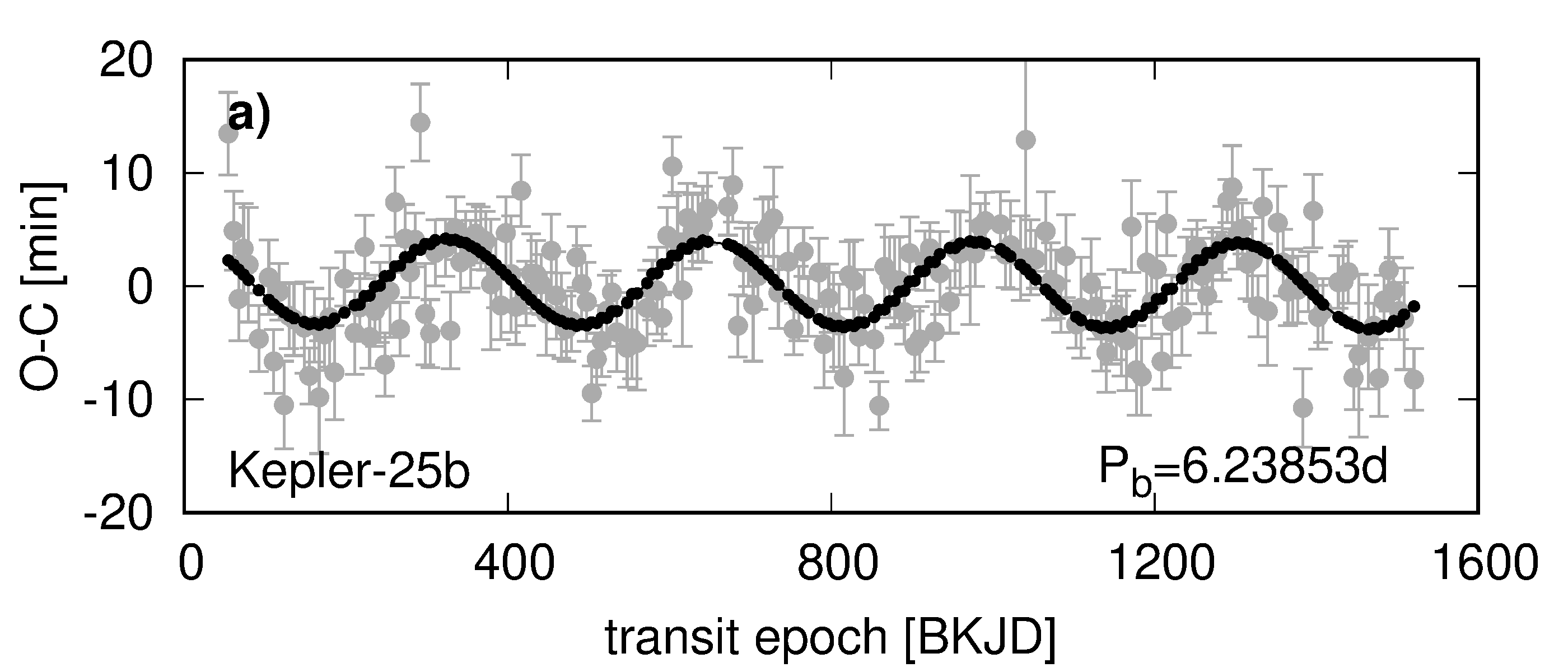}
}
\hbox{
\includegraphics[width=0.48\textwidth]{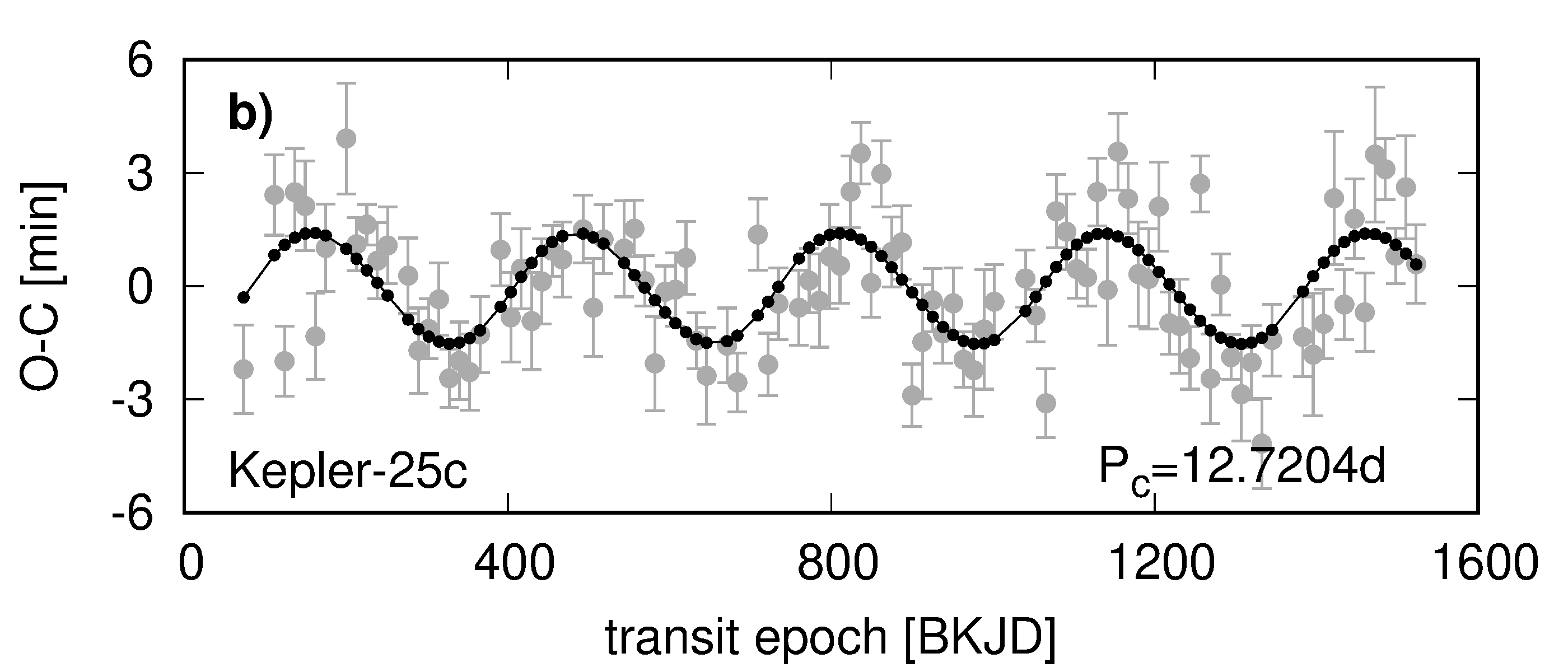}
}
}
}
\caption{TTV measurements (gray points with error bars) and the best-fitting model (black dots connected with lines in order to lead the eye).}
\label{fig:best_fit_OC}
\end{figure}

There is a clear minimum of $\Chi$ around $m_1 = 10.8\,\mE$ and $m_2 = 14.5\,\mE$. Those values agree with the results in \citep{Hadden2014}, where they used an analytic model of a near-resonant system and obtained $m_1 = (9.0 \pm 2.6)\,\mE$ and $m_2 = (14.3 \pm 4.5)\,\mE$. Formal confidence levels plotted with white curves indicate that the masses are constrained within a $\sim 1\,\mE$ uncertainty. The best-fitting model, that is presented together with the observations in Fig.~\ref{fig:best_fit_OC}, reconstruct the data satisfactorily well. Nevertheless, it is not perfect, as $\Chi \approx 1.59$ is greater than $1$ (what could be, in principle, explained by underestimated measurement uncertainties). The parameters of the best-fitting system are listed in Table~\ref{tab:params}. Formal $1-\sigma$ uncertainties are very small. One should keep in mind, though, that the orbital parameters are not free parameters of the model in the common sense. The results listed in the table has to be understood as the most likely values under the assumption that the configuration is periodic. The standard fitting procedure, in which the parameters are free, will be applied to the data further in this work.

{The parameters uncertainties listed in Tab.~\ref{tab:params} were computed with the uncertainty of the stellar mass was accounted for. The mass of the star is computed from the surface gravity (log$g$) and the radius of Kepler-25 listed in the catalogue in \citep{Rowe2015}. We obtain $m_0 = (1.19 \pm 0.05)\,\mS$. The stellar mass uncertainty enlarges uncertainties of the planets' masses (slightly) and of the semi-major axes (significantly). For the completeness of the errors estimates Tab.~\ref{tab:params} lists also the planet-to-star mass ratio as well as the Keplerian periods, computed from the semi-major axes with a help of the Kepler's third law.}

\begin{table}
\caption{The orbital elements of the best-fitting $(\Chi = 1.59)$ periodic configuration. The stellar mass {$m_0 = (1.19 \pm 0.05)\,\mS$} and the reference epoch $t_0 = 50.0$ (BKJD).}
\label{tab:params}
\begin{center}
\begin{tabular}{l c c}
\hline
parameter & planet~b & planet~c \\
\hline
{$m/m_0 [10^{-5}]$} & {$2.23 \pm 0.26$} & {$3.66 \pm 0.31$} \\
$m [\mE]$ & $10.8 \pm {1.1}$ & $14.5 \pm {1.3}$ \\
{$P [\mbox{d}]$} & {$6.23769(2)$} & {$12.7210(4)$} \\
$a [\au]$ & {$0.0703 \pm 0.0010$} & {$0.1130 \pm 0.0015$} \\
$e$ & $0.0014(1)$ & $0.00023(2)$ \\
$\varpi$ [deg] & $-57.639(1)$ & $101.31(33)$ \\
$\Mmean$ [deg] & $50.7358(33)$ & $-132.61(34)$ \\
\hline
\end{tabular}
\end{center}
\end{table}

\section{Can a non-resonant system mimic a periodic configuration?}

As we already mentioned, a periodic configuration is characterized by the (O-C)-signals in anti-phase. The observed system fulfils the criterion. Lets assume a sinusoidal model of TTV, i.e., (O-C)$_i = A_i \sin[(2\pi/\TOC) t + \Phi_i]$, where $\TOC$ is the period and $\Phi_i$ is the phase of the TTV signal of the $i$-th planet. By fitting the model to the data one obtains $A_1 = (3.8 \pm 0.4)\,$min, $A_2 = (1.6 \pm 0.2)\,$min, $\TOC = (325 \pm 5)\,$d and $\Deltaaph \equiv |(\Phi_1 - \Phi_2) - \pi| = (5 \pm 9)\,$deg. The latter quantity measures the deviation of the signals from the anti-phase. [Naturally, $(\Phi_1 - \Phi_2) - \pi$ is kept in the range of $(-\pi, +\pi)$.] For the periodic configuration $\Deltaaph = 0$.

\begin{figure*}
\centerline{
\vbox{
\hbox{
\includegraphics[width=0.44\textwidth]{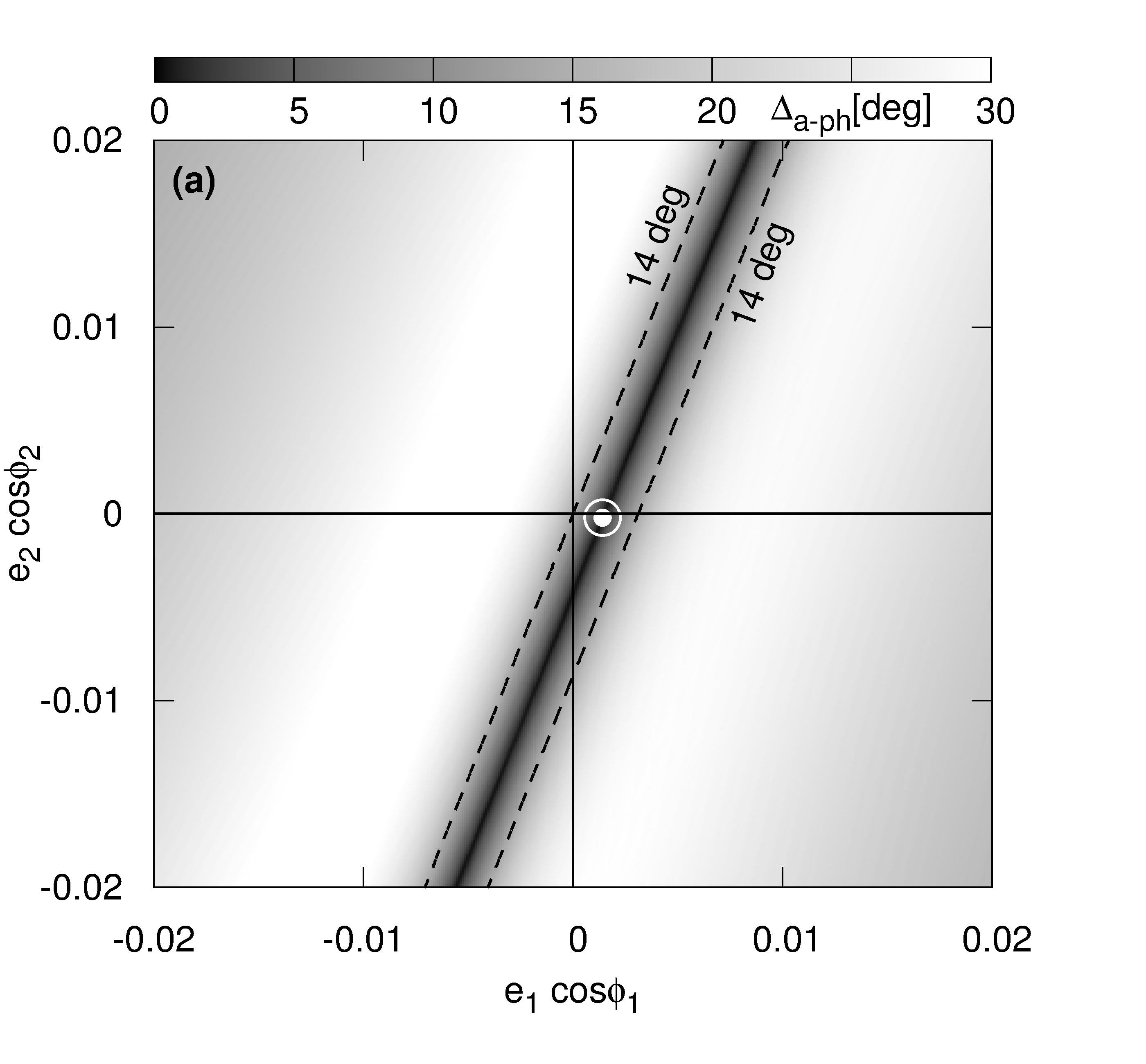}
\includegraphics[width=0.44\textwidth]{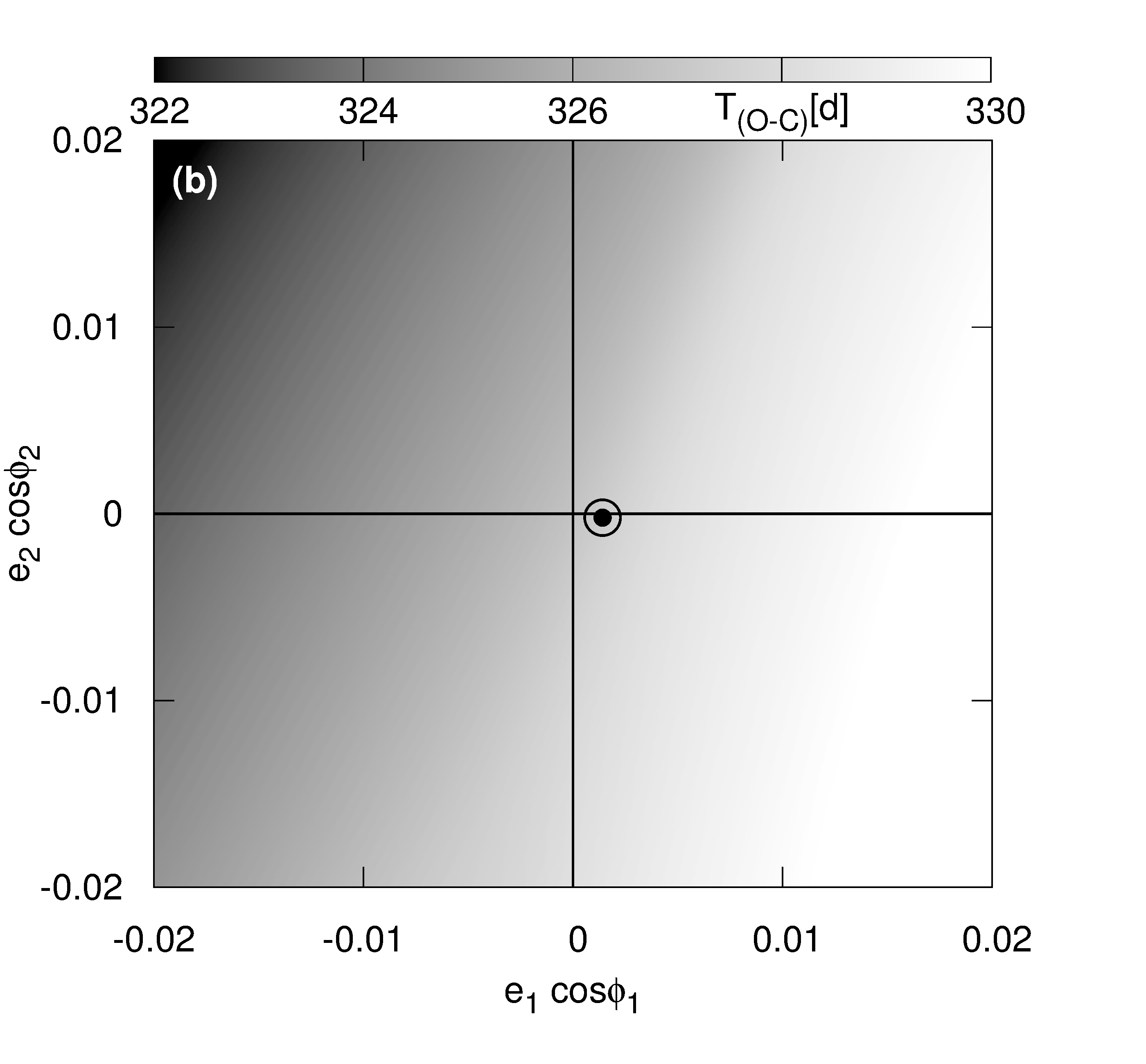}
}
\hbox{
\includegraphics[width=0.44\textwidth]{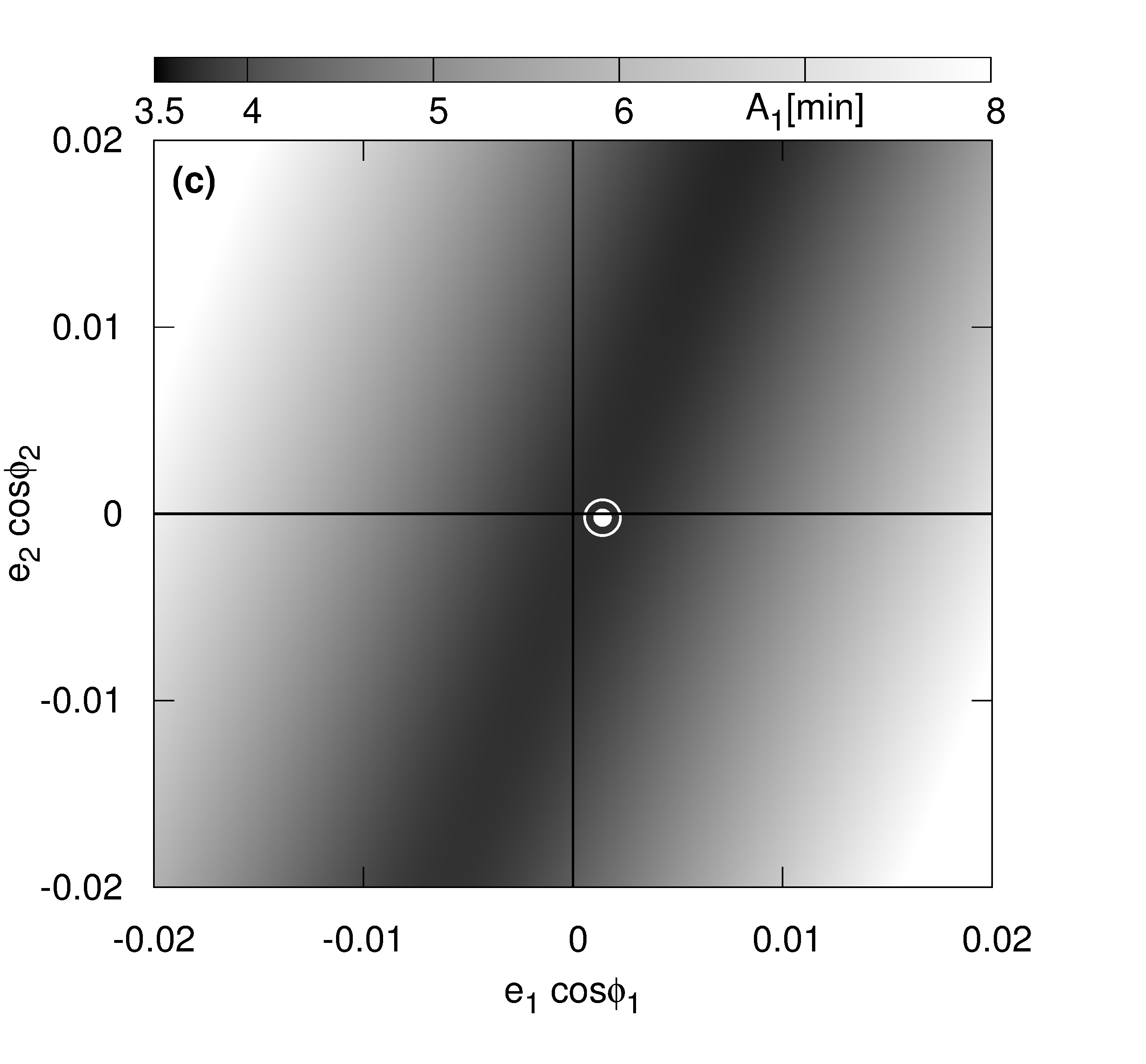}
\includegraphics[width=0.44\textwidth]{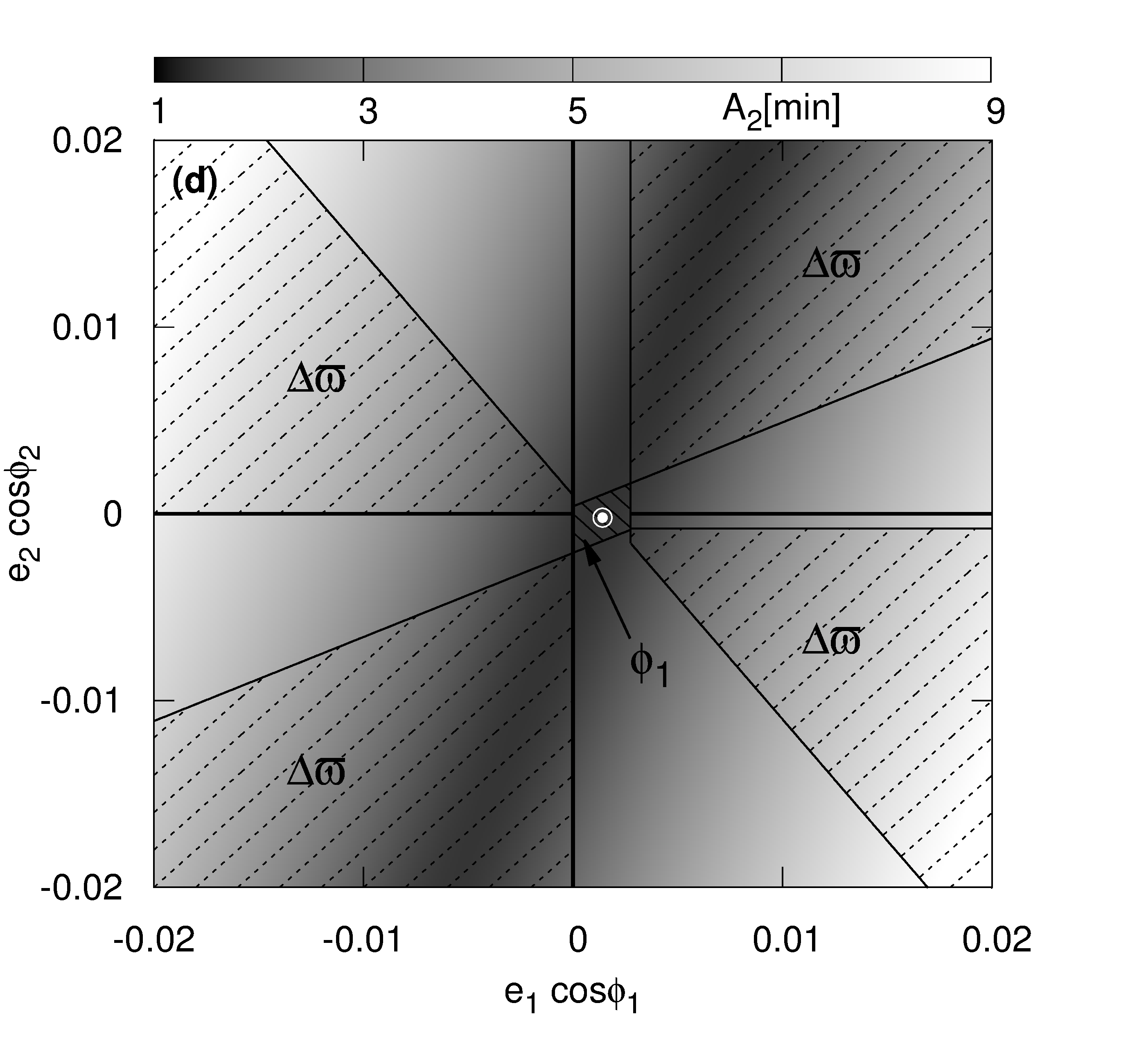}
}
}
}
\caption{Panel~(a): $(e_1 \cos\phi_1, e_2 \cos\phi_2)-$scan of the deviation from the anti-phase of the (O-C) synthetic signals, $\Delta_{\idm{a-ph}}$. Masses $m_1 = 10.8\,\mE, m_2 = 14.5\,\mE$ and semi-major axes $a_1 = 0.070274\,\au, a_2 = 0.1129985\,\au$ are fixed for the whole plane. Each quarter was obtained for different combinations of the angles $(\phi_1, \phi_2) = (0, 0), (0, \pi), (\pi, 0)$~or $(\pi, \pi)$. The white circle symbol points the position of the periodic configuration that fits the Kepler-25 TTVs. Panel~(b): an analogous plot presenting $\TOC$ as a function of the $(x, y)$. The bottom panels illustrates the scans of $A_1(x,y)$ and $A_2(x,y)$. Panel~(d) presents an additional information on the regions in which $\Delta\varpi$ and $\phi_1$ oscillate (dashed areas).}
\label{fig:e1e2_scan}
\end{figure*}

One may ask what is the value of $\Deltaaph$ for a system far from periodic. In general, we expect that $\Deltaaph$ is a function of initial orbital elements. Figure~\ref{fig:e1e2_scan}a presents a scan of this quantity in a plane of the eccentricities for a fixed period ratio and four representative pairs of the resonant angles $(\phi_1, \phi_2)$. Each quarter of the plane represents a different combination of the angles, i.e., $(0,0), (0,\pi), (\pi,0)~$and $(\pi,\pi)$ (lets enumerate those quarters by I, II, III and~IV, respectively). That means that there are two quarters with $\Delta\varpi=0$ and two with $\Delta\varpi=\pi$.

A given representative combination of $(\phi_1, \phi_2)$ can be achieved by two different combinations of $(\Mmean_1, \Mmean_2, \varpi_1, \varpi_2)$. The choices for particular quarters are dictated by the continuity requirement at the $x$ and $y$ axes of the plane, where $x = e_1 \cos\phi_1$ and $y = e_2 \cos\phi_2$. We chose for quarter~I $(\Mmean_1, \Mmean_2, \varpi_1, \varpi_2) = (0, 0, \pi, \pi)$, for quarter~II $(0, \pi, \pi, 0)$, for quarter~III $(\pi, 0, 0, \pi)$ and for quarter~IV $(\pi, \pi, 0, 0)$. The period ratio equals $2.039$ and the planets' masses equal the best-fitting values of $m_1 = 10.8\,\mE$ and $m_2 = 14.5\,\mE$. The white symbol points the position of the periodic configuration. The values of $\Delta_{\idm{a-ph}}(x,y)$ (that are coded with a shade of gray) are computed by integrating the N-body equations of motion for a particular initial system, computing the series of TTs (in a window of $4.3~$yr, as it is for the Kepler-25 system) and computing the phases of TTV signals for both planets. 

There is clearly a line corresponding to $\Delta_{\idm{a-ph}}=0$ (given by $y = 2.78722 x - 0.00412482$) going through the periodic system, what means that other configurations with different $(x,y)$ also produce TTV signals in anti-phase. By looking at this kind of plot alone, one cannot distinguish between the periodic and non-periodic (non-resonant) systems lying at the black line in Fig.~\ref{fig:e1e2_scan}a. Dashed lines indicate a value of $14$~degrees, what is the maximal $\Delta_{\idm{a-ph}}$ for the observed system.

Remaining panels of Fig.~\ref{fig:e1e2_scan} show scans of other quantities. Panel~(b) presents the results for $\TOC$. The period of the TTV signal does not change significantly over the plane. For the whole plane $\TOC$ agrees with the value of the Kepler-25 system, i.e., $(325 \pm 5)~$days. Panels~(c) and~(d) illustrate the results for the semi-amplitudes of the TTVs. The line for which $\Delta_{\idm{a-ph}}=0$ corresponds to minima of $A_1$ and $A_2$. At this line both $A_1$ and $A_2$ equal the values of the best-fitting model. In panel~(d) an additional information is given. Dashed areas indicate regions of oscillations of $\Delta\varpi$ and $\phi_1$. The latter encompasses a small area around the periodic configuration. The region in which the second resonant angle $\phi_2$ librates is smaller than the size of the white dot pointing the position of the periodic system. The areas of $\Delta\varpi$-oscillations are wide and exist for both oscillations centres $0$ and $\pi$. They correspond to non-resonant dynamics and represent the two modes of secular oscillations \citep[e.g.,][]{Michtchenko2004}. That means that a non-resonant system can mimic the periodic configuration.

\begin{figure}
\centerline{
\vbox{
\hbox{
\includegraphics[width=0.48\textwidth]{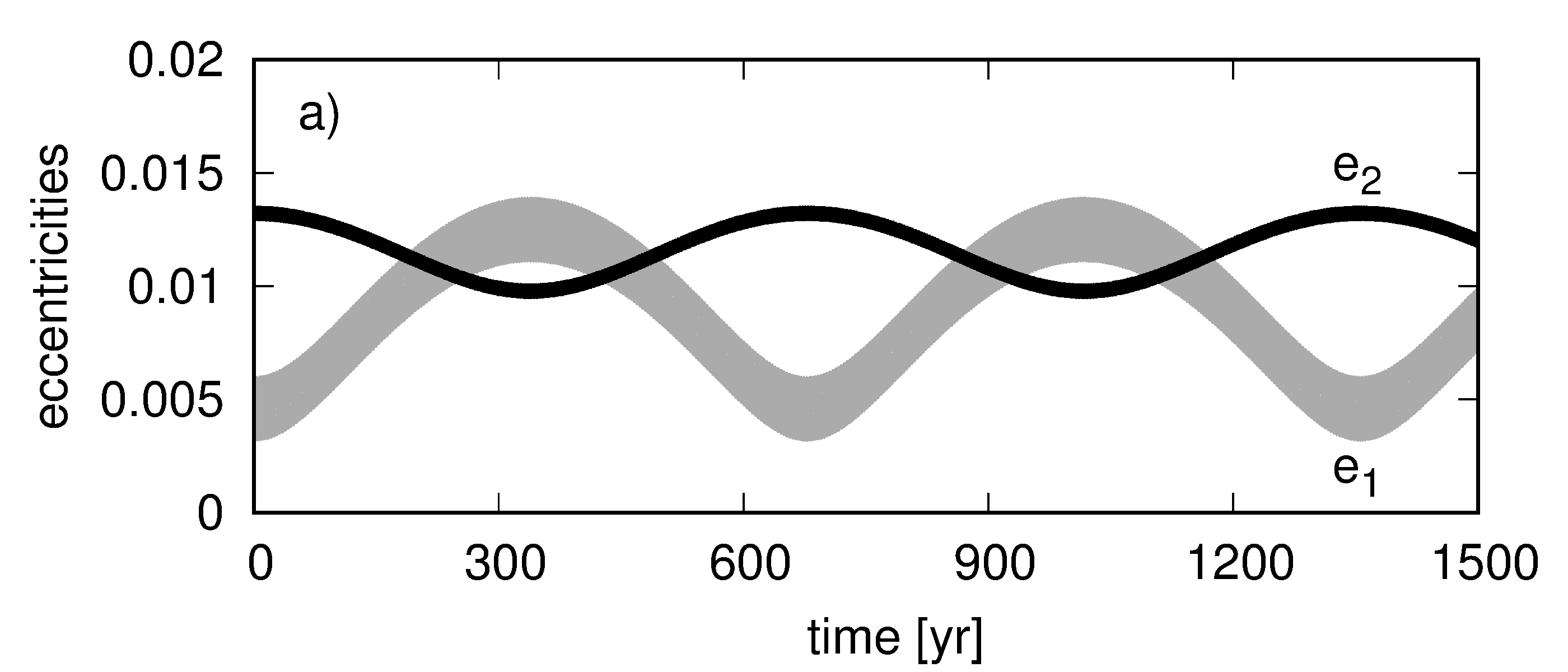}
}
\hbox{
\includegraphics[width=0.48\textwidth]{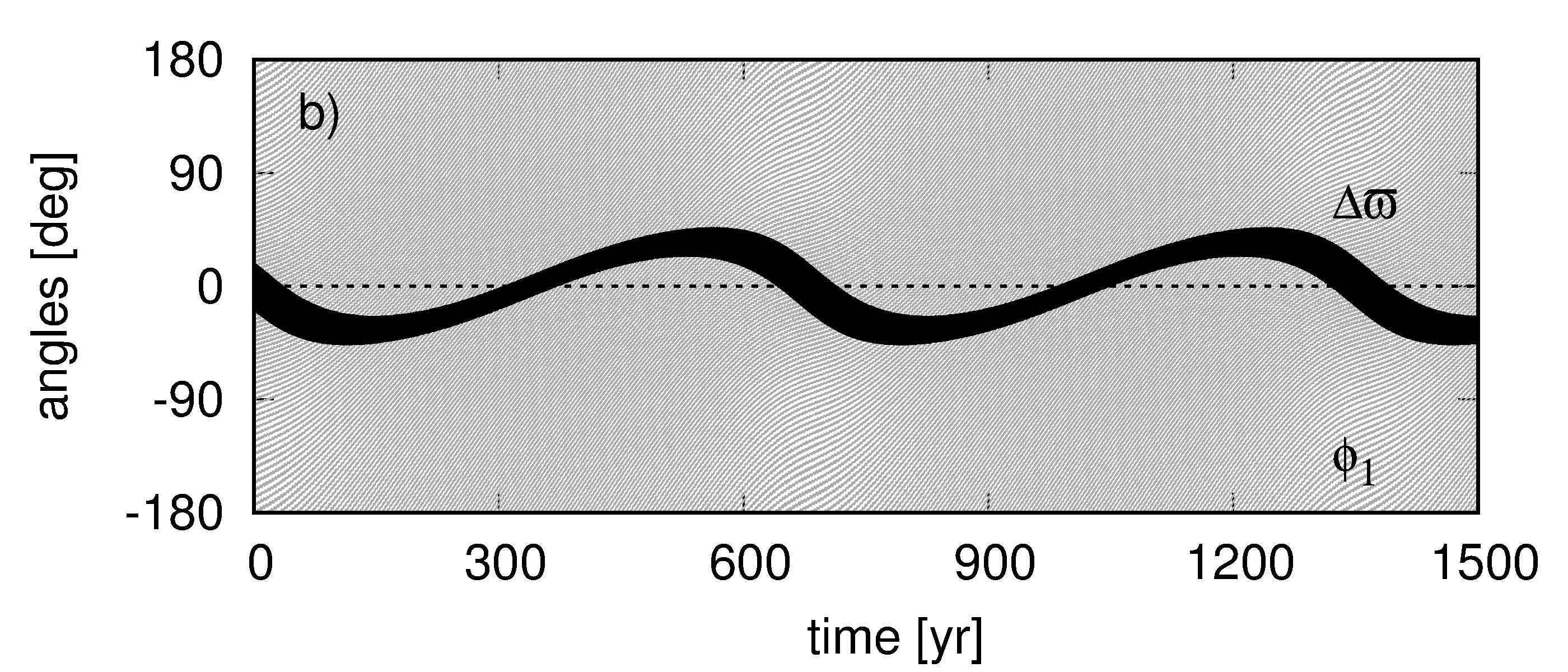}
}
}
}
\caption{Evolution of the eccentricities and the angles of an example initial configuration $e_1 = 0.006$, $e_2 = 0.013$, $P_2/P_1 = 2.039$, $\varpi_i = \Mmean_i = 0, i=1,2$.}
\label{fig:example_probability}
\end{figure}

\subsection{Probability that a non-resonant system mimics a periodic configuration}

A periodic configuration does not evolve in long time-scales (apart from a uniform rotation of the system as a whole), therefore, regardless the epoch in which we observe the system the (O-C)-diagram looks the same. On contrary, a non-periodic configuration evolves in the secular time-scale. In order to illustrate that we chose an initial system that lies in a vicinity of the $(\Delta_{\idm{a-ph}}=0)$-line, i.e., $e_1 = 0.006, e_2 = 0.013$ and all the angles equal $0$ (the point lies in quarter~I). The evolution of this example configuration is shown in Fig.~\ref{fig:example_probability}. The eccentricities as well as $\Delta\varpi$ vary in $\sim 680~$yr secular time-scale. The resonant angle $\phi_1$ rotates much faster. 

\begin{figure}
\centerline{
\vbox{
\includegraphics[width=0.48\textwidth]{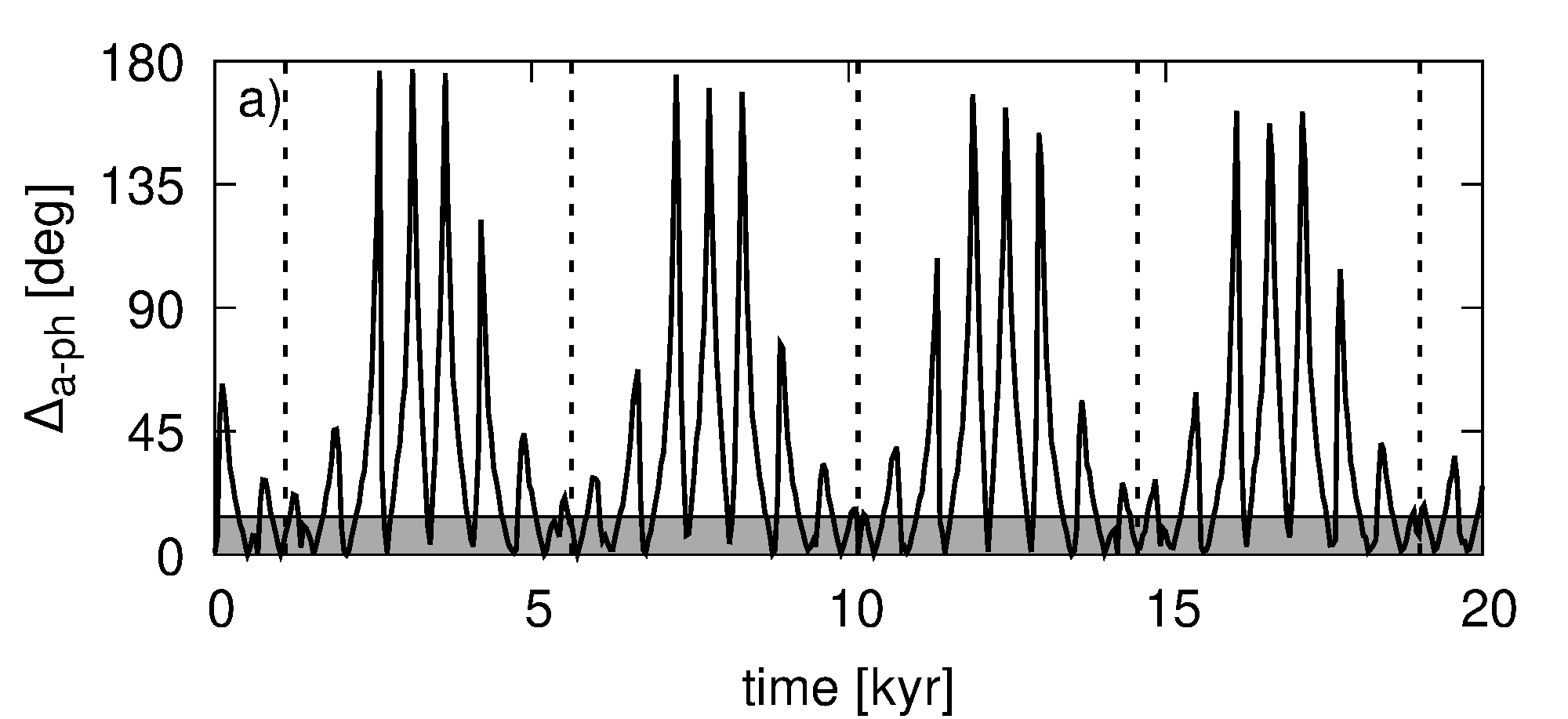}
\includegraphics[width=0.48\textwidth]{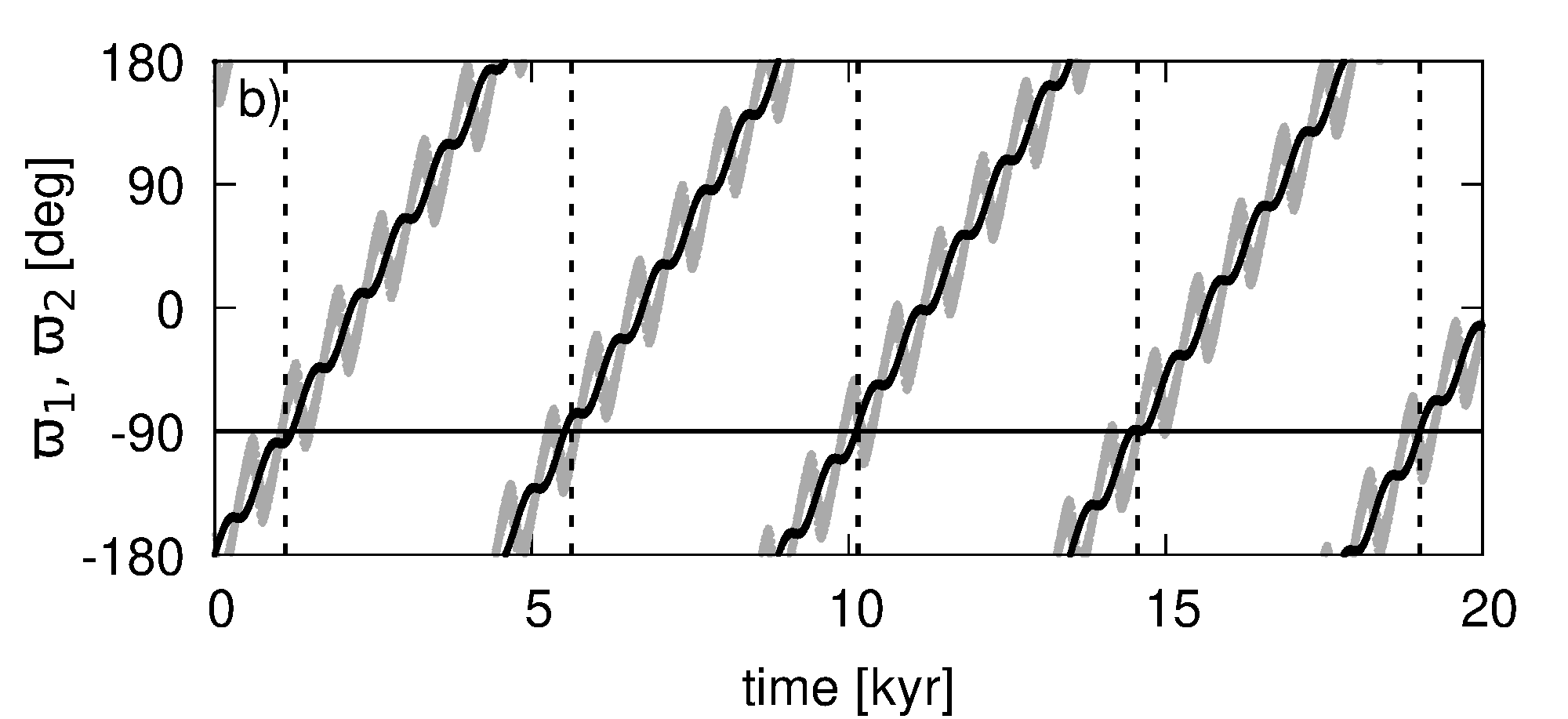}
}
}
\caption{Panel~(a): Variation of $\Deltaaph$ for the example initial configuration $e_1 = 0.006$, $e_2 = 0.013$, $P_2/P_1 = 2.039$, $\varpi_1 = \varpi_2 = \Mmean_1 = \Mmean_2 = 0$, whose orbital elements evolution was illustrated in Fig.~\ref{fig:example_probability}. Panel~(b): The evolution of the longitudes of pericentres in time for the same initial configuration (grey and black colours denote $\varpi_1$ and $\varpi_2$, respectively). Vertical dashed lines in both plots indicate epochs for which $\varpi_1 \approx \varpi_2 \approx -\pi/2$, that correspond to minima of $\sim 5\,$kyr-modulation of $\Deltaaph$.}
\label{fig:example_probability2}
\end{figure}

The evolution of the system results in the variation of $\Deltaaph$, what is illustrated in
Fig.~\ref{fig:example_probability2}a (note the wider time-window with respect to Fig.~\ref{fig:example_probability}). We observe the variation in $\sim 680~$yr period (reaching $\Deltaaph \sim 0$) as well as a longer-period modulation ($\sim 5\,$kyr), that corresponds to the rotation of the system as a whole (see Fig.~\ref{fig:example_probability2}b for the evolution of individual values of $\varpi_1, \varpi_2$). We observed that the long-period modulation of $\Deltaaph$ has minimal amplitude for $\varpi_1 \approx \varpi_2 \approx -\pi/2$. Therefore, a system that is not periodic may look like one (in a sense of $\Deltaaph \sim 0$) for particular orientations of the orbits as well as for particular phases of the secular evolution.

A shaded area in Fig.~\ref{fig:example_probability2}a denotes the $14~$degree limit. There are epochs in the evolution of this system in which the TTV signals can be even in phase, i.e., $\Delta_{\idm{a-ph}} = \pi$. As a result, only for some part of the time the system, that is not periodic, looks like a periodic configuration. One can compute the probability that $\Delta_{\idm{a-ph}}<14~$deg (denoted with $p_1$) by dividing the amount of time in which $\Delta_{\idm{a-ph}}<14~$deg by the whole time of the integration. 
Because $\Deltaaph$ depends on the two characteristics, i.e., on the phase in the secular modulation of the eccentricities and the spatial orientation of the system, instead of integrating a given initial configuration for a very long time, we integrate the system for the time that equals the secular period of $\sim 680\,$yr and rotate the configuration within the whole range of $360~$degrees. The probability is computed by evaluating $\Deltaaph$ every $20\,$yr of the integration and by rotating the system at each epoch with an increment of $20\,$deg. That makes $34 \times 18 = 612$ (O-C) diagrams for each initial configuration to be tested. A number of diagrams for which $\Deltaaph<14~$degrees divided by $612$ defines $p_1$.  
For the system considered $p_1 \approx 40$~per~cent. In general, the probability $p_1$ is a function of $(x,y)$.

\begin{figure*}
\centerline{
\vbox{
\hbox{
\includegraphics[width=0.44\textwidth]{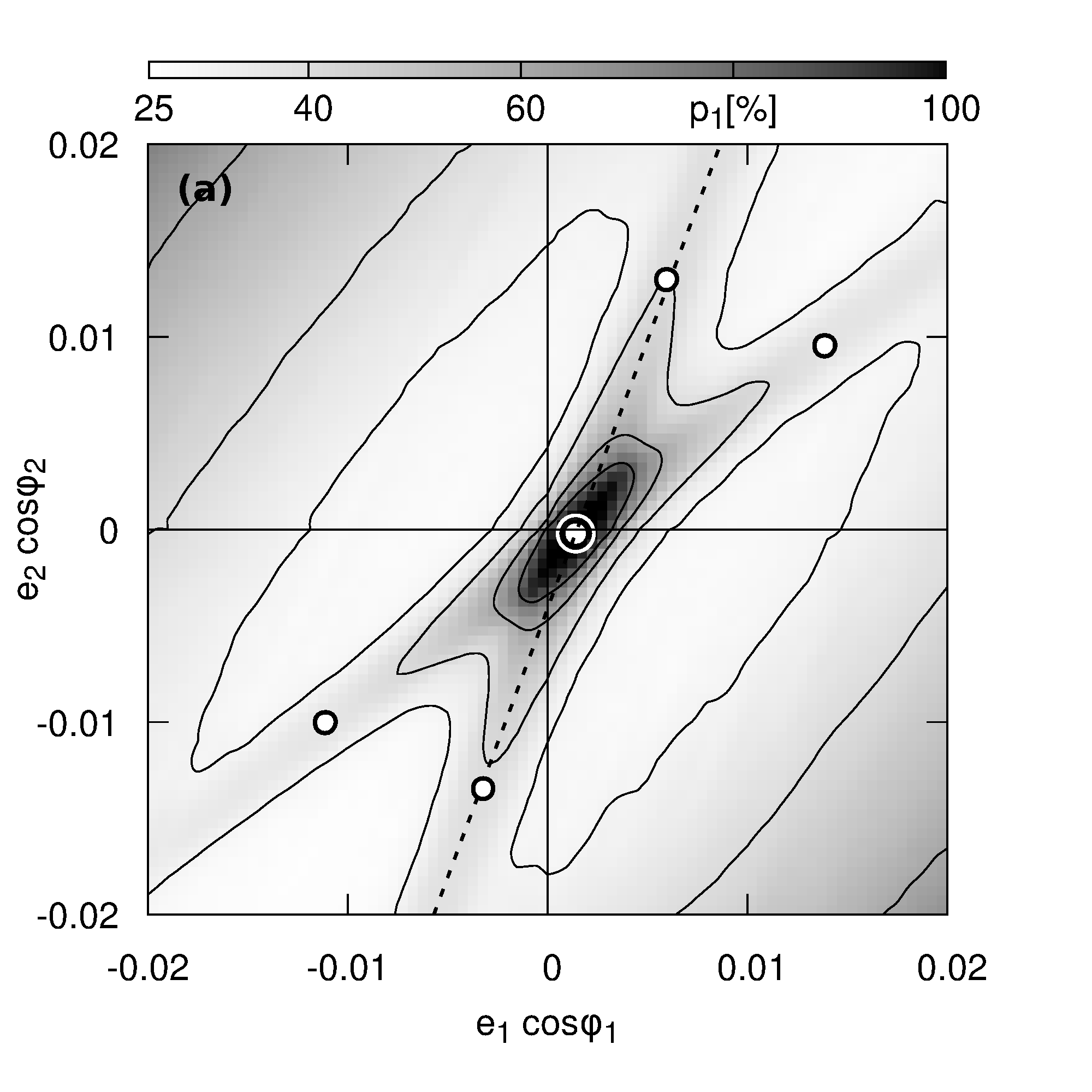}
\includegraphics[width=0.44\textwidth]{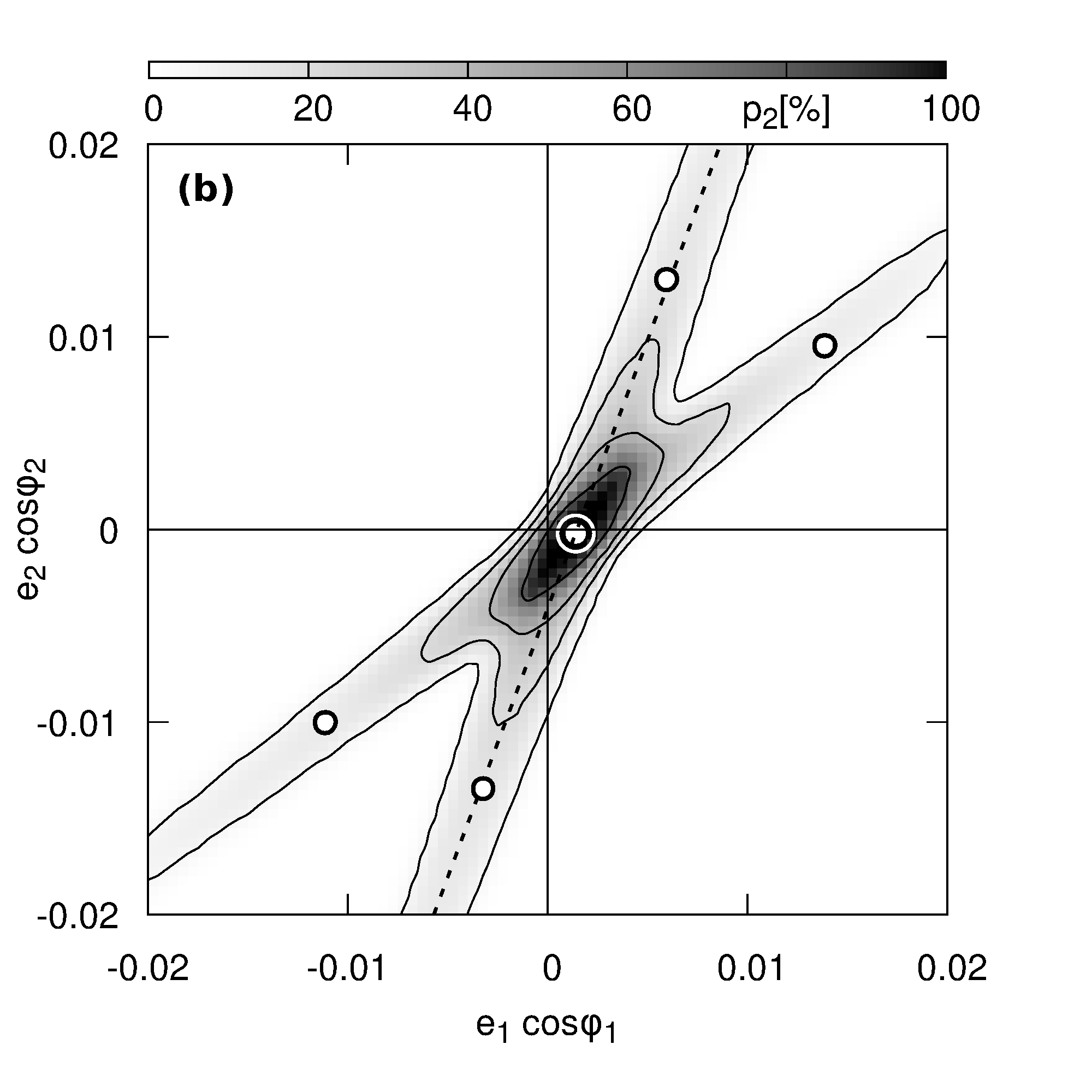}
}
}
}
\caption{Panel~(a): A scan of the probability $p_1$ presented at the representative plane. The black circles indicate the points in which the example configuration, whose evolution was illustrated in Figs.~\ref{fig:example_probability} and~\ref{fig:example_probability2}, intersects the plane. The big white symbol close to the centre represents the periodic configuration, while the dashed line corresponds to $\Delta_{\idm{a-ph}}=0$ at the representative plane. Panel~(b): A scan of the probability $p_2$ presented in the same manner as in panel~(a).}
\label{fig:scan_probability}
\end{figure*}

The $(x,y)-$scan of the probability $p_1$ is presented in Fig.~\ref{fig:scan_probability}a. Naturally, for the periodic system and in its vicinity $p_1 = 100~$per~cent. The region of relatively high $p$ forms an X-shape structure at the plane. That stems from the fact that a given point at the plane has three counterparts, as a given configuration, in general, intersects the representative plane in four points during the evolution. The positions of the intersections for the example configuration with $e_1 = 0.006$ and $e_2 = 0.013$ ($\phi_1 = \phi_2 = 0$) are marked with black circles. 

One can see that for higher eccentricities, especially in quarters~II and~III, $p_1$ is relatively high. We computed the scan in a wider range of $e < 0.1$, and observed that $p_1$ can reach even $100~$per~cent. Panels~(b), (c) and~(d) of Fig.~\ref{fig:e1e2_scan} show that not only $\Deltaaph$ depends on $(x,y)$, also $\TOC$ and the amplitudes $A_1, A_2$ are functions of the eccentricities and the resonant angles. Moreover, those quantities depend on the phase in the secular evolution as well as the orientation of the system. Therefore, a given system with high $p_1$ can be, in general, characterized by $\TOC, A_1, A_2$ very different from the values of the observed system. Such a system cannot be considered as consistent with the observations. 

\begin{figure}
\centerline{
\includegraphics[width=0.48\textwidth]{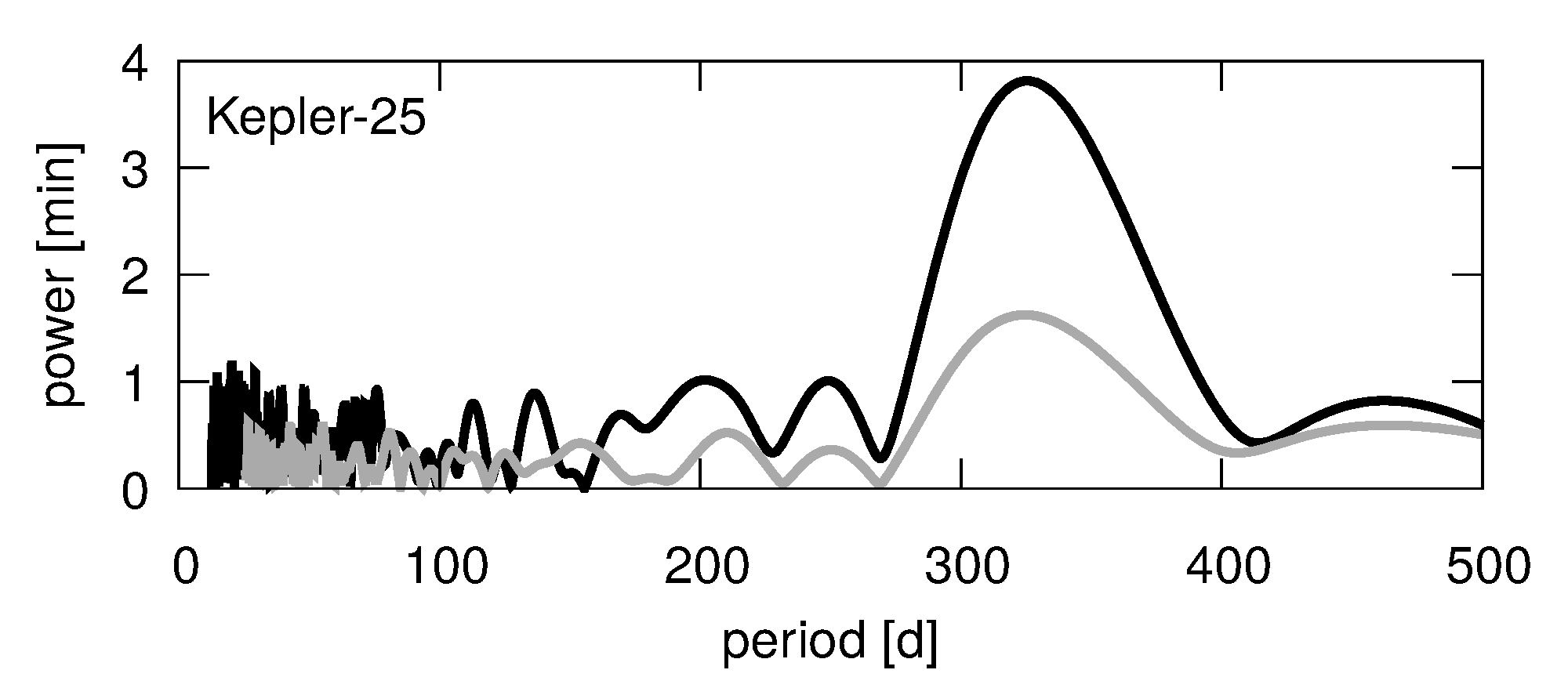}
}
\caption{The Lomb-Scargle periodograms of the (O-C)-signals of Kepler-25 (black and grey curves are for the inner and the outer planets, respectively).}
\label{fig:k25_lsp}
\end{figure}

Another characteristic of the (O-C)-signal that needs to be considered is whether or not there is a second mode in it. Figure~\ref{fig:k25_lsp} illustrates the Lomb-Scargle periodograms of the (O-C)-signals of the Kepler-25 system. There is no secondary peak higher than $\sim 0.27$ and $\sim 0.33$ of the primary peaks heights (for planets~b and~c, respectively).
Therefore, we require that for the synthetic systems considered as consistent with the observations the secondary-to-primary peak ratio is below the limits given above.

\begin{figure*}
\centerline{
\vbox{
\hbox{
\includegraphics[width=0.48\textwidth]{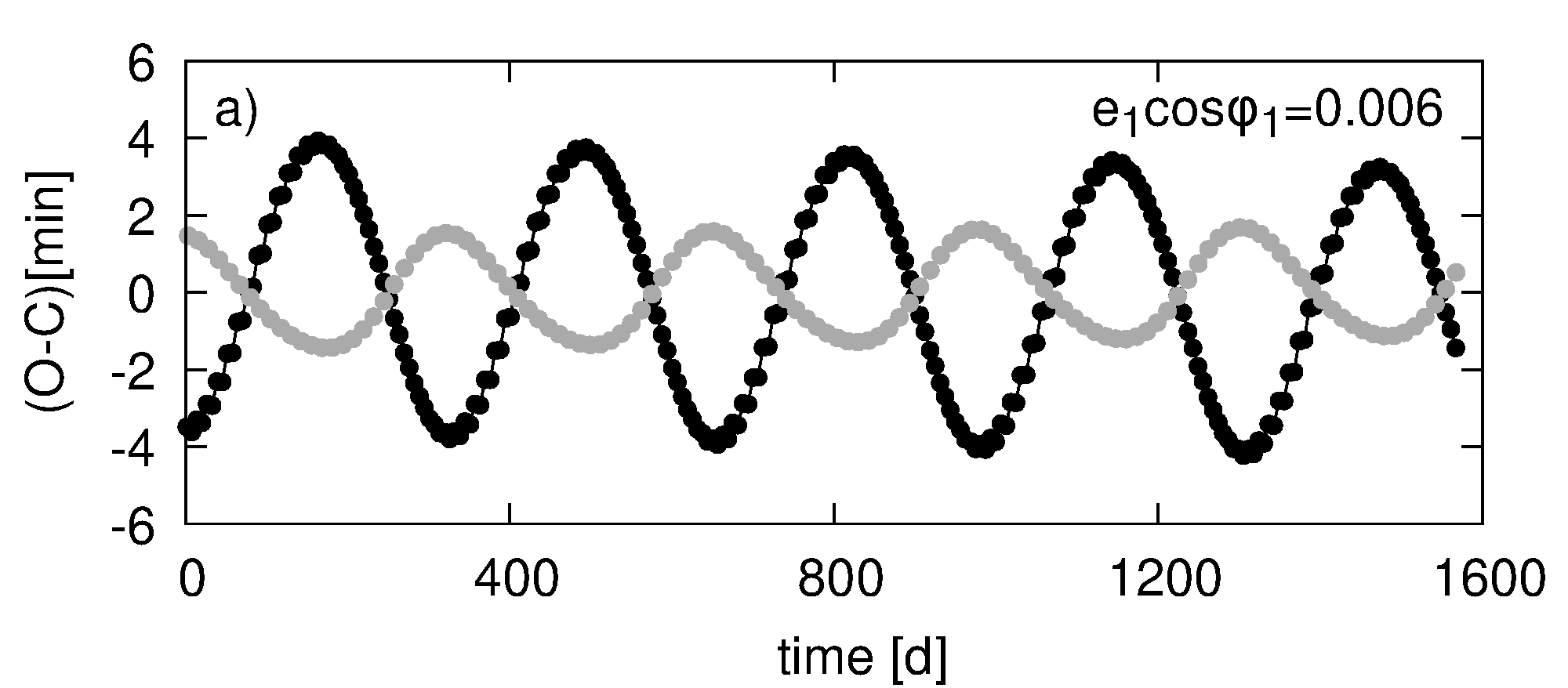}
\includegraphics[width=0.48\textwidth]{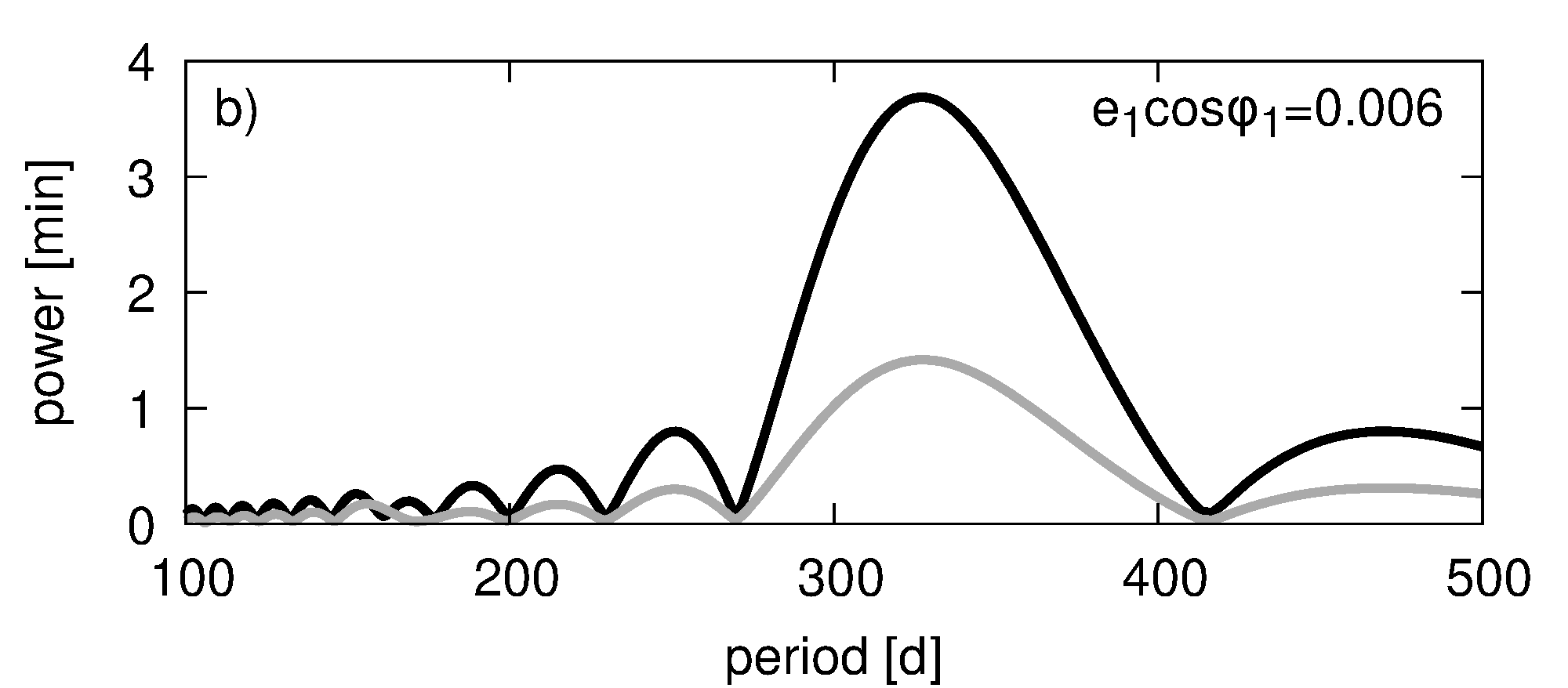}
}
\hbox{
\includegraphics[width=0.48\textwidth]{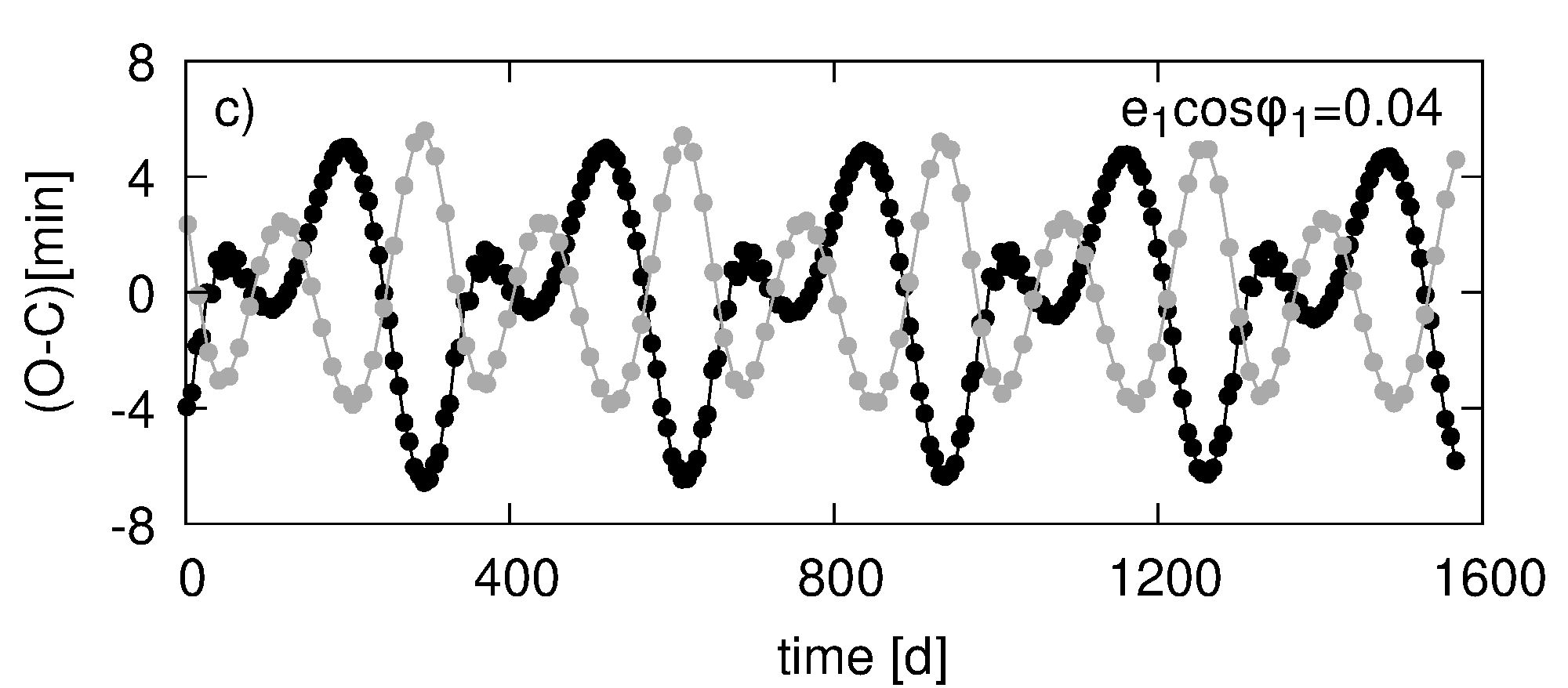}
\includegraphics[width=0.48\textwidth]{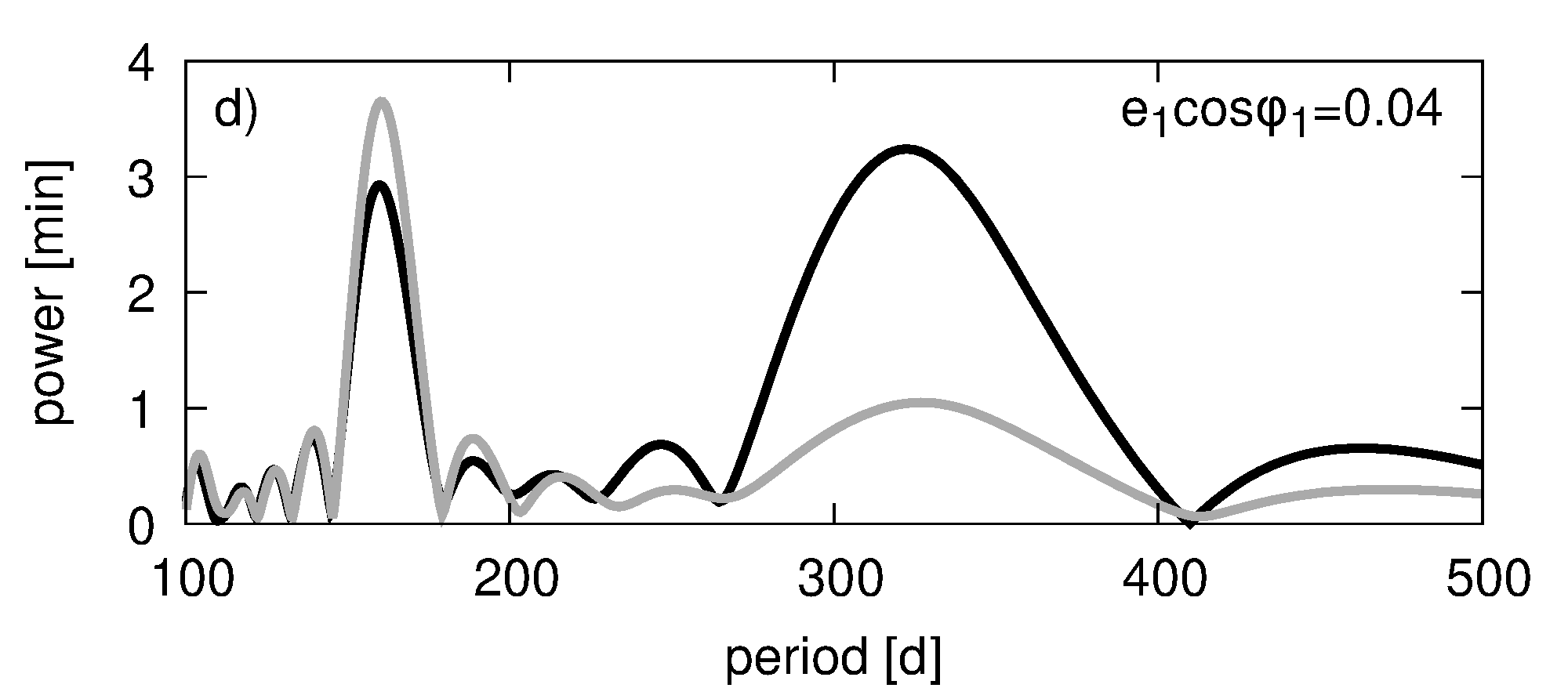}
}
}
}
\caption{The left-hand column: the (O-C)-plots for two configurations with $(e_1, e_2)$ equal to $(0.006, 0.013), (0.04, 0.107)$, for the top and bottom rows respectively. For each of them $\Delta_{\idm{a-ph}} \sim 0$. Both the systems are chosen from quarter~I of the representative plane. The right-hand column: the Lomb-Scargle periodograms for the signals presented in the left-hand column. Black and grey curves correspond to the inner and the outer planets' signals, respectively.}
\label{fig:ttv_lsp}
\end{figure*}

Figure~\ref{fig:ttv_lsp} presents the (O-C)-signals as well as the Lomb-Scargle periodograms for two configurations that have $\Deltaaph \sim 0$ (see the caption of this figure for the parameters). The system with $e_1 = 0.006$ has a uni-modal (O-C)-diagram, as it should be for a close-to-periodic configuration. The system with higher eccentricities is characterized with bi-modal (O-C)-signals. Therefore, this kind of configuration is not consistent with the observations.

In order to incorporate those criteria into the probability that a given configuration from the $(x, y)-$plane can be the real configuration of Kepler-25, we defined another quantity, $p_2$, which is a probability that a given system has $\Deltaaph < 14~$deg and also that $A_1 \in (2.6, 5.0)\,$min, $A_2 = (1.0, 2.2)\,$min, $\TOC = (320, 330)\,$d and that the second peaks in the periodograms of both signals are smaller than the values given above. The ranges of the allowed amplitudes correspond to the $3-\sigma$ confidence levels for the values of the observed signal. We widened the range because the amplitudes depend on the planets' masses, that are fixed for the plane at the values constrained with the periodic configuration assumption, that is not fulfilled for the whole plane.

Figure~\ref{fig:scan_probability}(b) shows the scan of $p_2$ in the same manner as in panel~(a). Clearly, the only configurations that might correspond to the observed system are confined to the X-shape structure around the periodic configuration. Still, the probability that a non-resonant low-eccentric configuration lying in a vicinity of the line given by $y = 2.78722 x - 0.00412482$ can mimic the periodic configuration is relatively high. We computed a 1-dimensional scan of $p_2$ along the $(\Deltaaph = 0)$-line (not shown) and found that for $e_1 \gtrsim 0.01$ (and $e_2 \gtrsim 0.027$) the probability is already very low, $p_2 \lesssim 5\,$per~cent, while for $e_1 \gtrsim 0.02$ (and $e_2 \gtrsim 0.054$) $p_2 = 0$.

\section{The MCMC analysis}

Before going to the MCMC (Markov Chain Monte Carlo) analysis of the TTVs we consider an influence of the third non-transiting planet in the system on the TTVs of the two transiting planets. \cite{Marcy2014} measured the radial velocities (RV) of the Kepler-25 and constrained the masses of planets~b and~c to be $m_1 = (9.6 \pm 4.2)\,\mE$ and $m_2 = (24.6 \pm 5.7)\,\mE$, that is roughly consistent with the model of the periodic configuration, although the mass of planet~c is slightly smaller in our work. They also found that apart from the two planets discovered by the \kepler{} mission \citep{Steffen2012}, there is a weak RV signal of the third, more distant companion.
The amplitude of the RV variation is relatively small and the precision of the measurements was not good enough to constrain the parameters of the third planet satisfactorily. There are two possible periods of planet~d reported, i.e., $(123 \pm 2)$ and $(93 \pm 2)$ days. The mass of the planet $m_3 = (89.9 \pm 13.7)\,\mE$ and the eccentricity $e_3 = 0.18$ {(the uncertainty of $e_3$ was not given)}. There is a possibility, though, that the third planet affects noticeably the TTVs of the two transiting planets~b and~c. We tried to verify the possibility by performing a following test.

{An expected result of an existence of the third planet in relatively wide orbit is an additional signal in (O-C) of the period equal to the orbital period of this planet and its harmonics \citep{Agol2005}. The perturbing planet should not change the phase nor the period of the (O-C) of the resonant pair. In order to verify that and to check the amplitudes of the additional (O-C)-modulation due to the outermost companion, we }
added the third planet into the model, constructed the (O-C)-diagrams for planets~b and~c and compared the diagrams with the ones obtained for the unperturbed two-planet system. {At first we chose the most likely values of $m_3$ and $e_3$.} For planet~b the difference is practically none, while for planet~c {the additional signal amplitude $\lesssim 0.1~$min}, depending on the assumed period of the third planet ($123$ or $93$~days) and a given angles $(\Mmean_3, \varpi_3) = (0, 0), (0, \pi), (\pi, 0)$ or~$(\pi, \pi)$. The $A_2$ variation is well below the TTVs uncertainties, moreover {neither} the period {nor the phase} of the (O-C)-signal {are} altered due to the third planet.
{Next, we increased the mass of the outermost companion to $89.9 + 13.7 = 103.6\,\mE$ and tried to find at what value of $e_3$ (when increasing above the most likely value of $0.18$) the amplitude of the additional (O-C)-modulation equals $\sim 0.5\,$min, that is the noise level for (O-C) of planet~c (see the Lomb-Scargle periodogram illustrated in Fig.~\ref{fig:k25_lsp}). We found that for $P_3 = 123~$days, the limiting $e_3 = 0.4$, while for $P_3 = 93~$days, the limiting $e_3 = 0.3$. For eccentricities higher than the limiting values the third planet would produce an additional signal in (O-C)-diagram of planet~c that is detectable. As we do not observe any additional periodicities in the (O-C)-diagrams, we conclude that the eccentricity $e_3$ is below the limiting values we found and the long-period companion does not need to be incorporated into the model.}

\subsection{The standard TTV modelling and the periodic configuration}

At present, the Bayesian inference is a de facto standard  for the analysis of the \kepler{} light-curves and the TTV measurements. A crucial step in this approach is to define correct priors for determining the posterior distribution of model parameters. This is particularly important for interpretation of TTV models which
tend to degenerate solutions characterised by strongly aligned orbits with moderate and large eccentricities \citep[e.g.,][]{Hadden2014,JontofHutter2016,MacDonald2016}. Therefore, the TTV fitting must be monitored whether or not its results depend on the adopted eccentricity priors, to avoid drawing incorrect conclusions on the orbital
archituecture of the studied planetary system. 

\cite{JontofHutter2016} assumed the Rayleigh distribution for the prior of the eccentricities and chose two different values of the Rayleigh parameter $\sigma_e = 0.1$ (wide prior, i.e., weak constraints on the eccentricities) and $\sigma_e = 0.02$ (narrow prior, i.e., stronger constraints). The posterior distributions of $\Delta\varpi$ for a few systems obtained for the two different values of $\sigma_e$ differ significantly one from another. For the wide prior the most likely are the aligned configurations, i.e., with $\Delta\varpi \sim 0$. When the prior is narrower, the peaks of the posterior distributions of $\Delta\varpi$ move towards $\pm \pi$, although for the value they used, $0.02$, {the maxima of the posterior distribution are shifted by only $20-50~$degrees with respect to $0$}, depending on the system. On may expect, that for even lower $\sigma_e$, the peaks could move to the $\pm \pi$ positions. 

We showed previously (Fig.~\ref{fig:scan_probability}) that the aligned non-resonant configurations can mimic the periodic configuration. The (O-C)-signals of such qualitatively different systems look the same, what may bring problems when the direct fitting approach is applied. The eccentricities of the periodic configuration for $P_2/P_1 = 2.039$ are very small, $\sim 0.0001 - 0.001$, therefore, finding such a system without strong a~priori constraints on the eccentricities may be difficult. We presume that non-periodic (aligned) configurations are favoured in the fitting procedure (with wide priors put on the eccentricities), regardless the real architecture of the system. 

In this section we aim to verify this hypothesis by running extensive MCMC sampling of the posterior distribution. We used the TTV model described, for instance, in our previous papers \citep{Gozdziewski2016,Migaszewski2017}. Similarly, we imposed the Gaussian priors with the mean equal to 0 and a prescribed standard deviation $\sigma$ set on the Poincar\'e parameters, $X \equiv e \cos\varpi$ and $Y \equiv e \sin\varpi$, rather than on the eccentricities. The Poincar\'e elements are the free parameters of the TTV model and encode the eccentricity and the longitude of pericentre, respectively. We sampled the posterior distribution with the {\tt emcee} package by \citep{ForemanMackey2013}, choosing up to 256,000 iterations made with 1024 ``walkers'' in a small hypercube in the parameter space. 

\begin{figure*}
\centerline{
\vbox{
\hbox{
\includegraphics[width=0.316\textwidth]{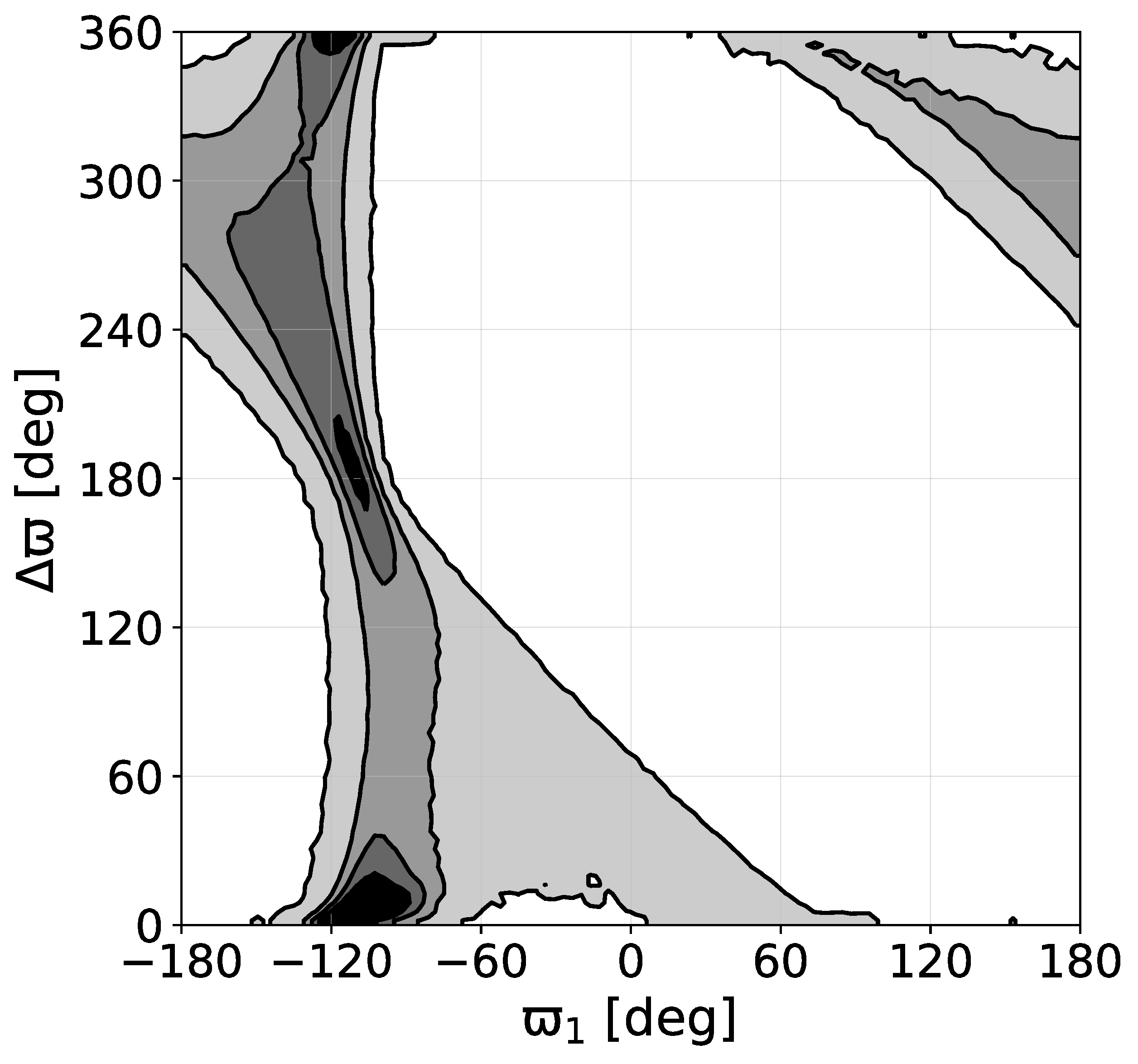}
\includegraphics[width=0.298\textwidth]{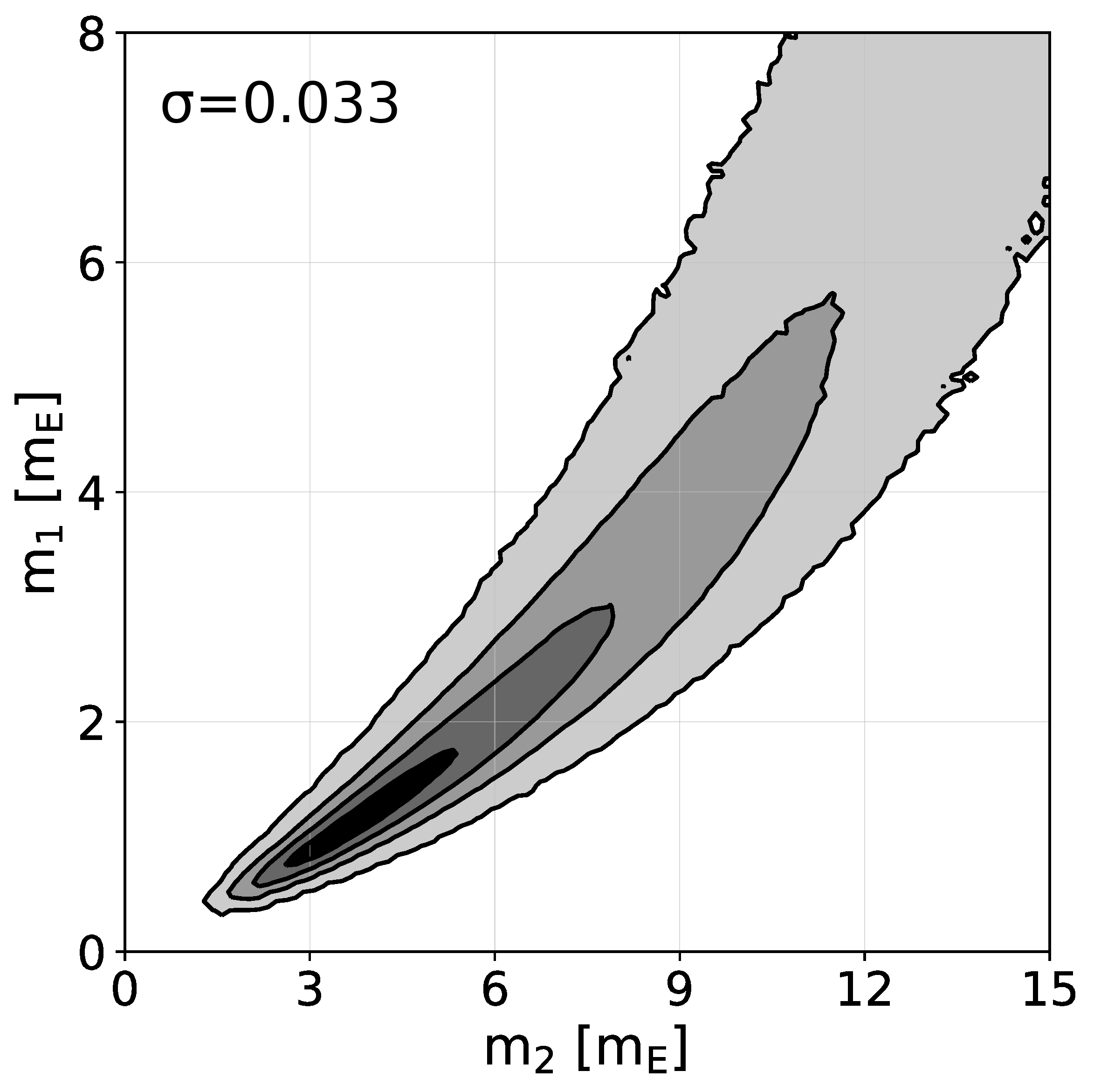}
\includegraphics[width=0.320\textwidth]{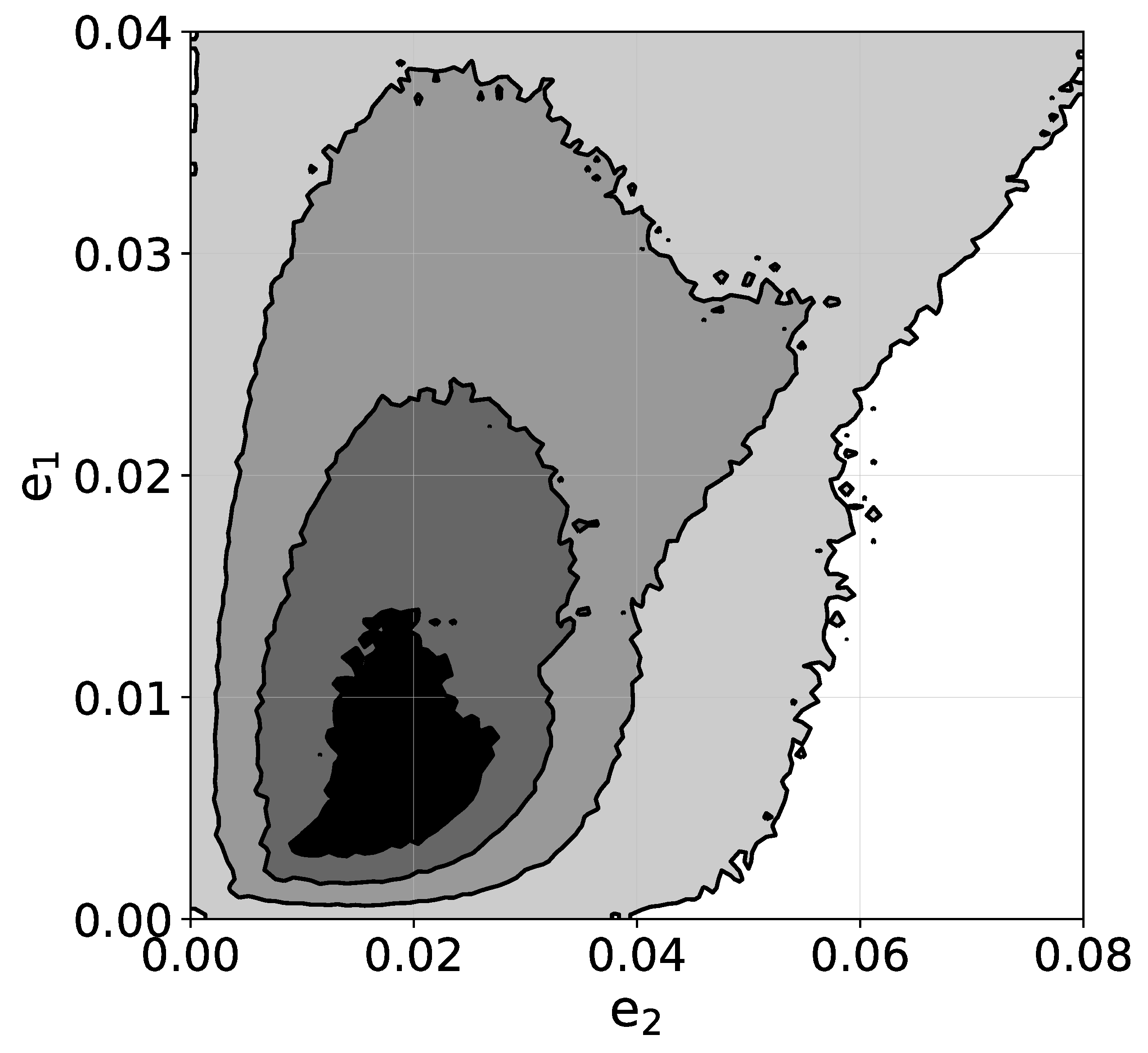}
}
\hbox{
\includegraphics[width=0.310\textwidth]{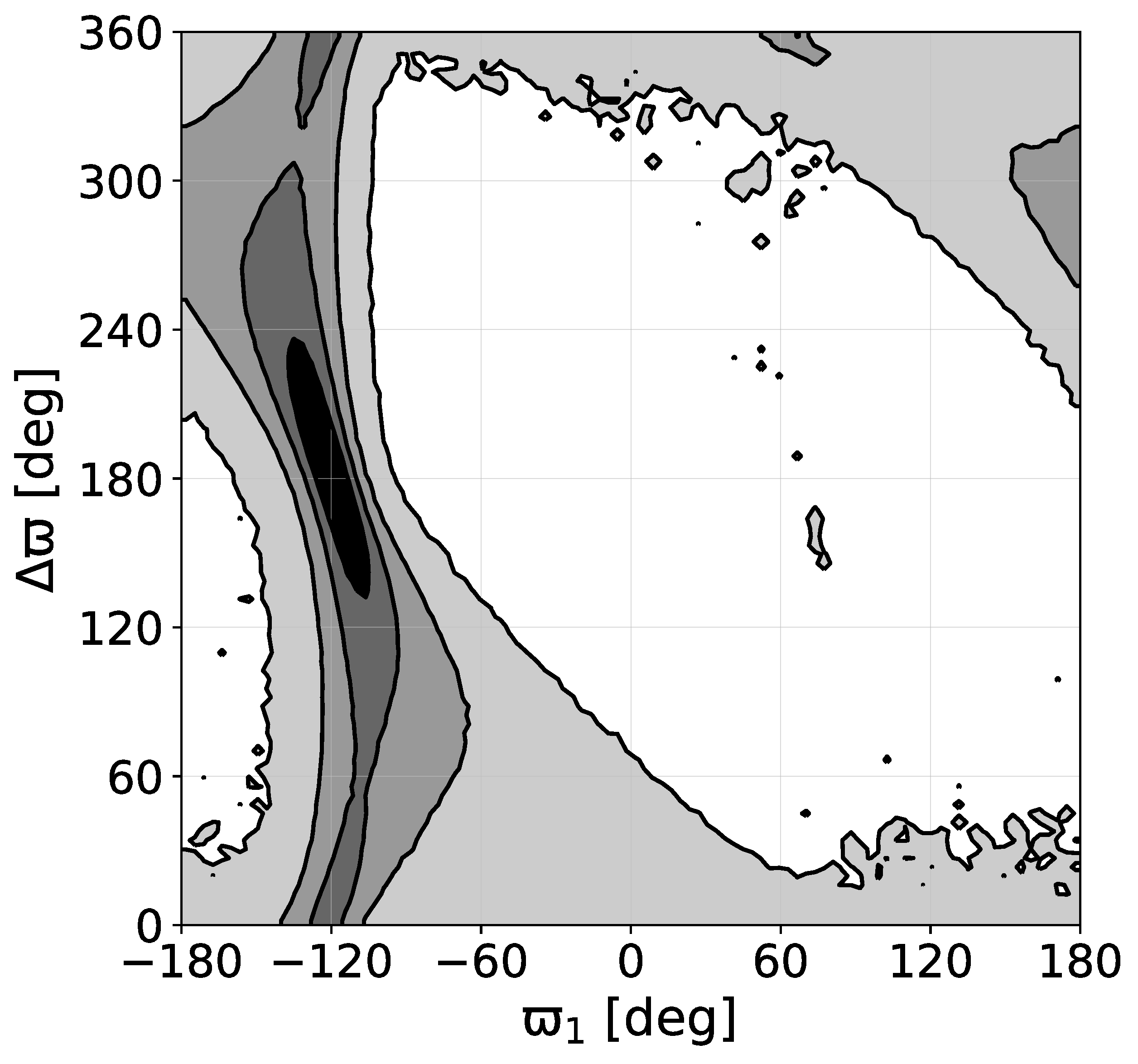}
\includegraphics[width=0.300\textwidth]{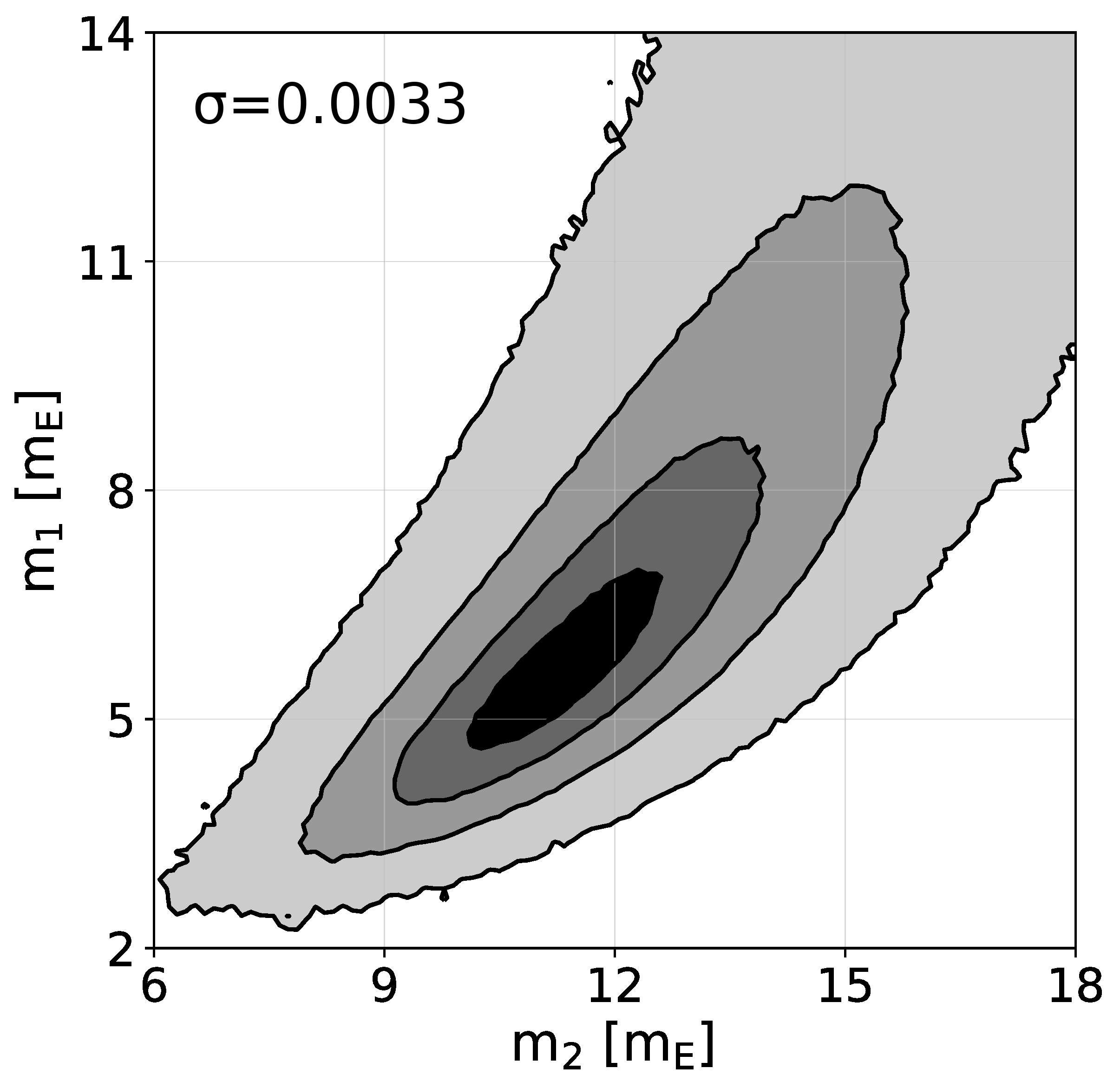}
\includegraphics[width=0.330\textwidth]{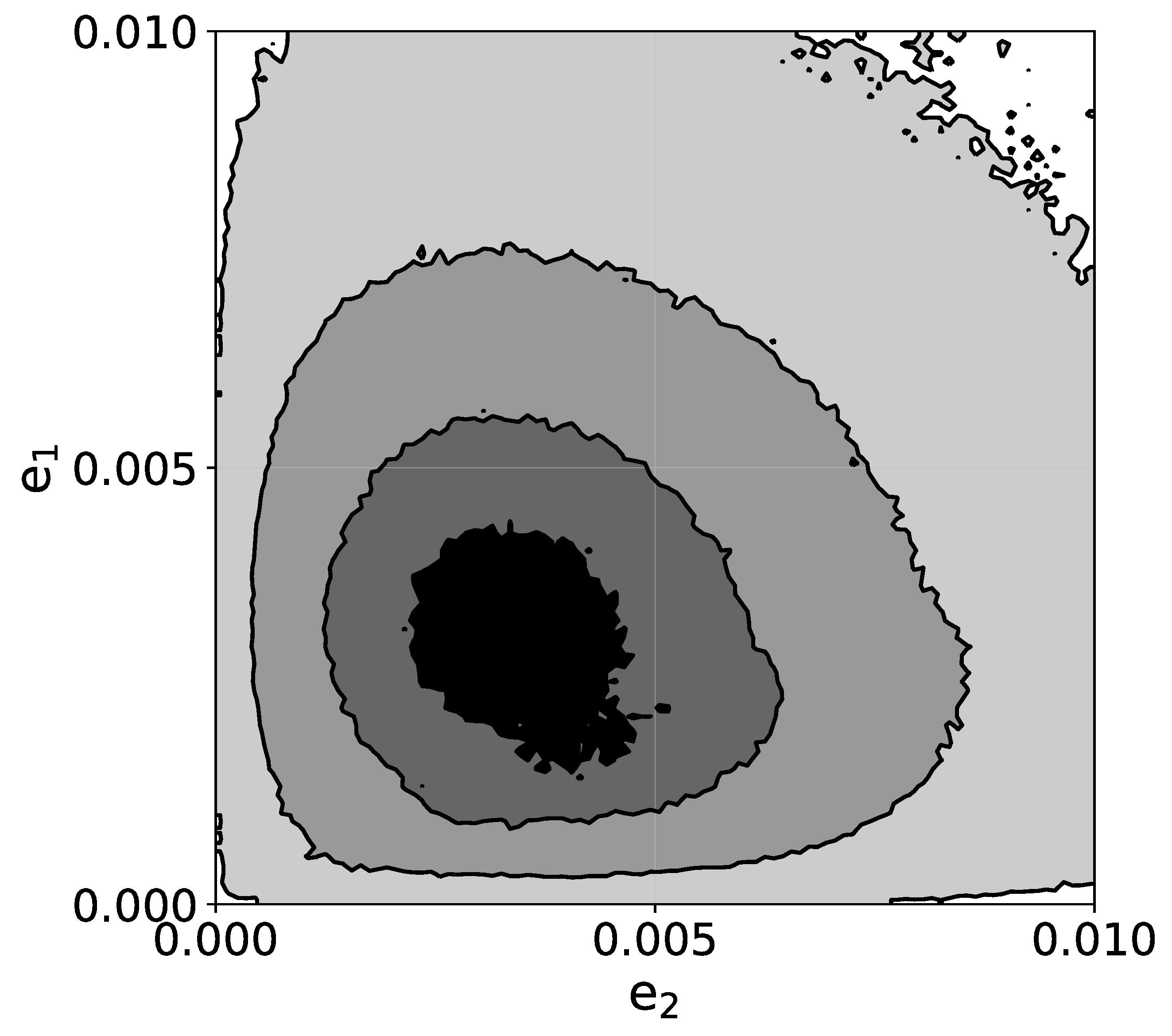}
}
\hbox{
\includegraphics[width=0.310\textwidth]{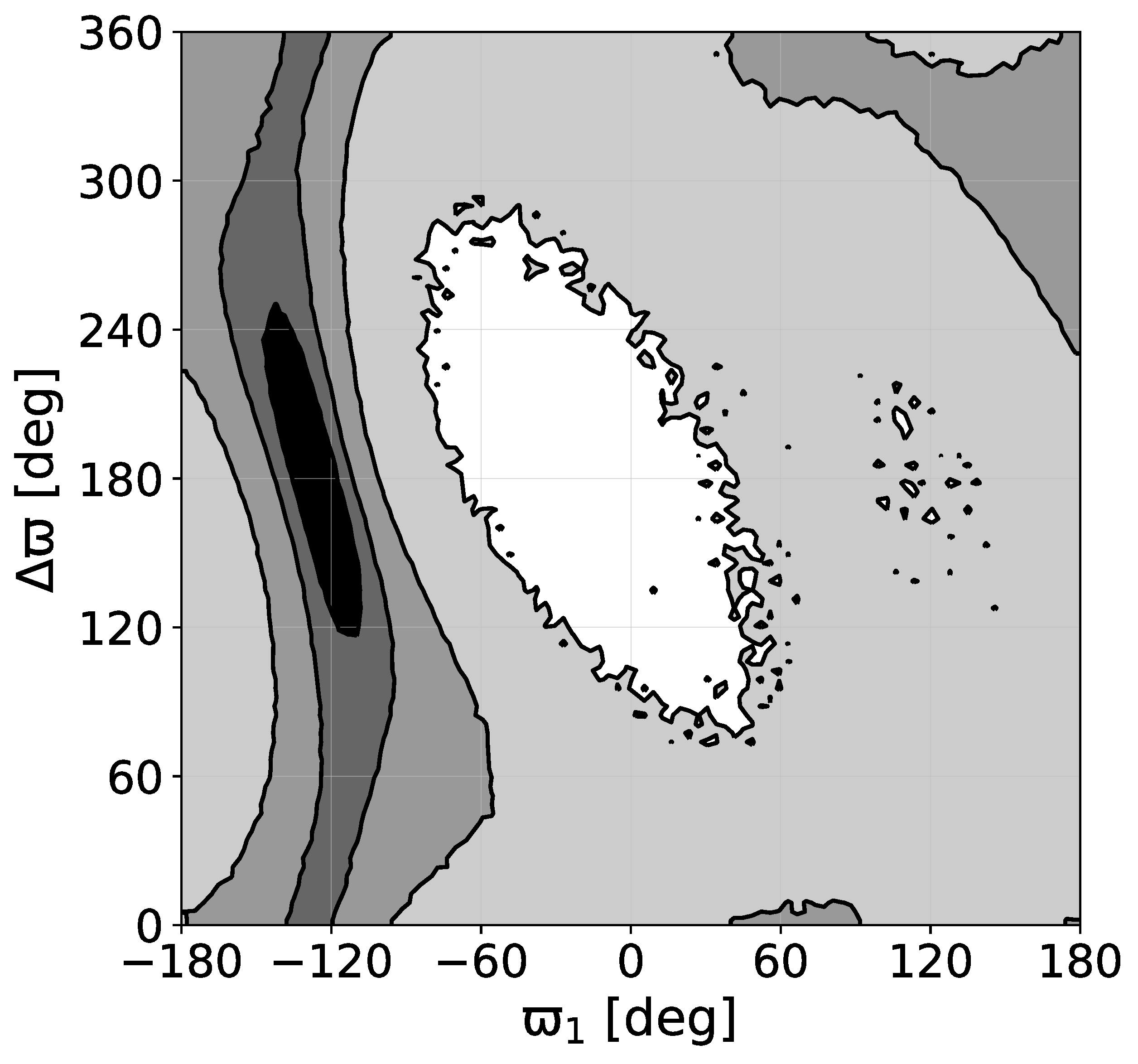}
\includegraphics[width=0.300\textwidth]{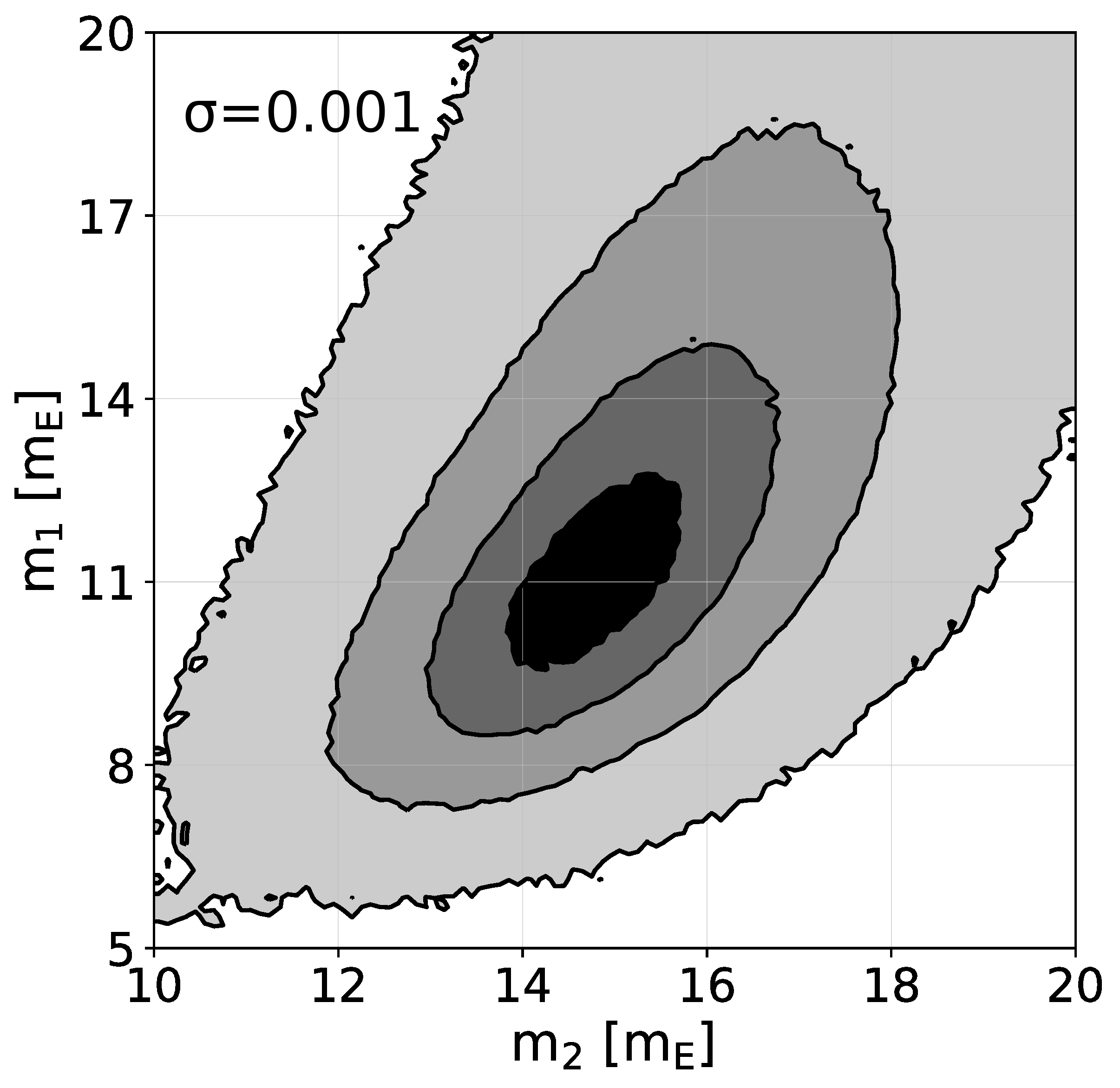}
\includegraphics[width=0.332\textwidth]{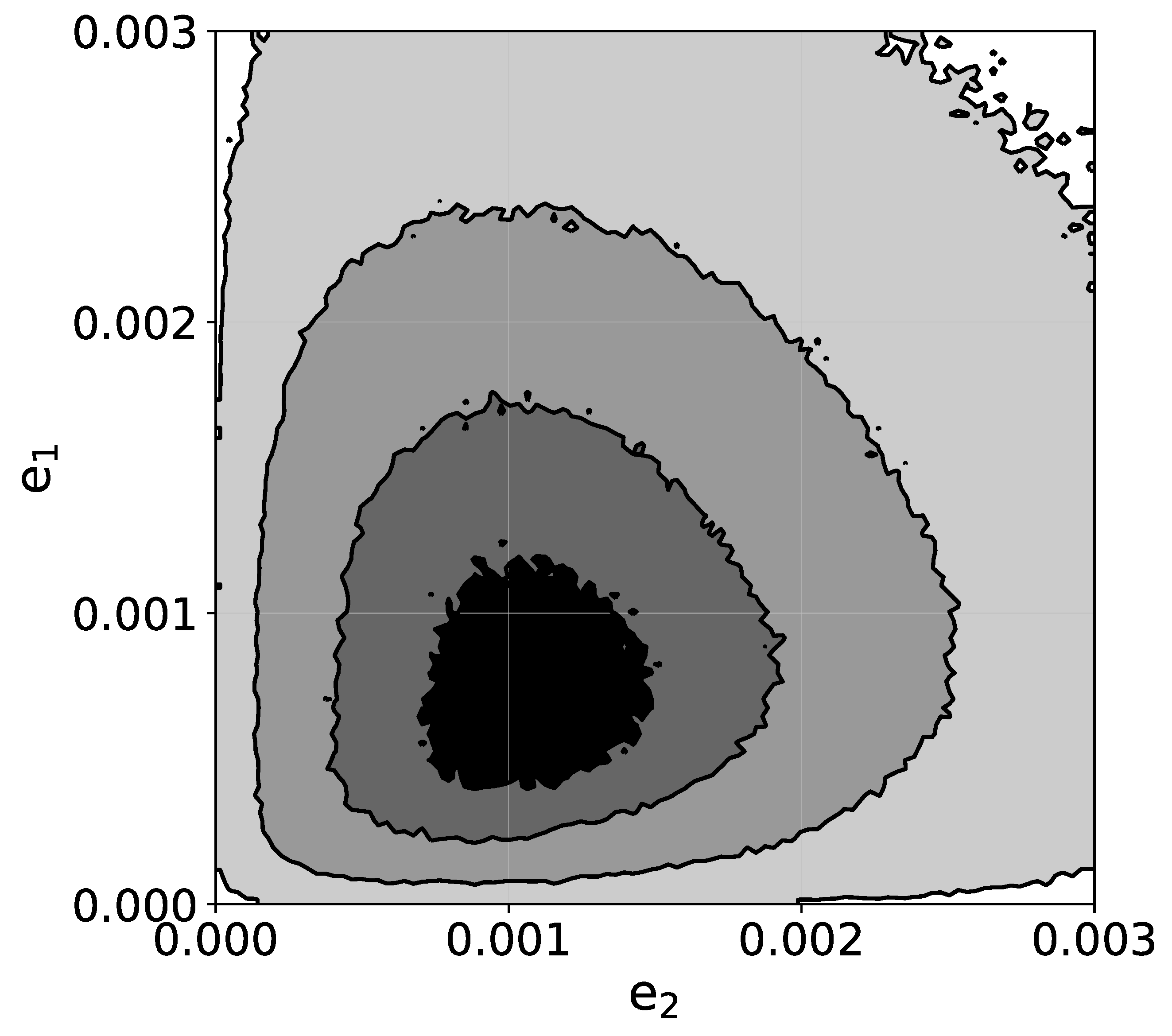}
}
}
}
\caption{
Two--dimensional projections of the posterior probability distribution
for the Kepler-25 system. Columns are for the eccentricities, 
masses (expressed in Earth masses), the longitude of pericentre ($\varpi_1$)
and the apsidal angle $\Delta\varpi$ (expressed in degrees), respectively.
Rows from the top to the bottom are for $\sigma=0.033$, $0.0033$ and $0.001$, respectively. {Notice that the axes ranges for the middle and the righ-hand columns are different for different sigma.}
Contours are plotted for the $14$-, $50$-th, $86$- and $99.9$ percentile of the MCMC samples derived from runs of 1024 walkers for 128,000 iterations
each; for $\sigma=0.033$ the number of walkers was increased to 2048
in order to address the two-modal posterior distribution.
}
\label{fig:mcmc.new}
\end{figure*}

We did a series of the MCMC experiments for 
$\sigma \in [0.00033, 0.1]$.
The results, selected for representative values of $\sigma=0.033,0.0033,0.001$, are illustrated in Fig.~\ref{fig:mcmc.new}. Each row presents the results for one of the values of $\sigma$ (given in the middle-column panels). The left-hand column shows the posterior probability distribution at the plane of $(\varpi_1, \Delta\varpi)$, the middle column is for the $(m_1, m_2)$-plane, while the right-hand column -- for $(e_1, e_2)$-plane.
These results confirm our predictions. Indeed, for high $\sigma$ aligned configurations are preferred, while lower $\sigma$ leads to anti-aligned orbits. Although for high $\sigma$ the anti-aligned systems are also allowed, they are less likely than the aligned ones. Another general observation is an existence of a correlation between $\sigma$ and the masses and eccentricities, i.e., higher $\sigma$ means lower masses and higher eccentricities.

The top row of Fig.~\ref{fig:mcmc.new} illustrates the results for $\sigma = 0.033$, which may be representative for high $\sigma$, as for even higher values the qualitative picture is the same. The posterior probability distribution is bi-modal, as both aligned and anti-aligned systems are possible. The aligned orbits are, as already mentioned, more likely. For higher $\sigma$ the disproportion between the modes is even more significant (the corresponding plots are not shown). The bi-modality is seen also at the eccentricities plane (right-hand column), although it is not that clear as for the $(\varpi_1, \Delta\varpi)$-distribution. The most likely masses are very low, i.e., $m_1 \approx 1\,\mE$, $m_2 \approx 4.5\,\mE$, which gives low densities for both the planets, $\sim 0.3\,$g/cm$^3$.

The second row from the top of Fig.~\ref{fig:mcmc.new} shows the posterior probability distribution obtained for $\sigma = 0.0033$. The anti-aligned mode becomes more significant than the aligned one. The masses increase and the eccentricities decrease when compared to the previous example. The masses increases by a factor of $6$ and $2.5$, for the inner and the outer planet, respectively. That leads to the densities, respectively, $\sim 1.8\,$g/cm$^3$ and $\sim 0.68\,$g/cm$^3$. Clearly, different prior information on the eccentricities leads to qualitatively different both dynamical structure of the system and possible physical composition of the planets. 

\begin{table}
\caption{The orbital elements of the best-fitting $(\Chi = 1.532)$ configuration from the MCMC analysis for $\sigma = 0.001$. The stellar mass {$m_0 = (1.19 \pm 0.05)\,\mS$} and the reference epoch $t_0 = 50.0$ (BKJD).}
\label{tab:params_mcmc}
\begin{center}
\begin{tabular}{l c c}
\hline
parameter & planet~b & planet~c \\
\hline
{$m/m_0 [10^{-5}]$} & {$3.11^{+0.94}_{-0.66}$} & {$3.89^{+0.43}_{-0.41}$}\\
$m [\mE]$ & {$12.3^{+3.8}_{-2.7}$} & {$15.4^{+1.8}_{-1.7}$} \\
{$P [\mbox{d}]$} & {$6.23768(6)$} & {$12.7210(9)$} \\
$a [\au]$ & {$0.0703 \pm 0.0010$} & {$0.1130 \pm 0.0015$} \\
$e$ & $0.0010^{+0.0008}_{-0.0006}$ & $0.0005^{+0.0008}_{-0.0005}$ \\
$\varpi$ [deg] & $-131^{+26}_{-43}$ & $58^{+182}_{-75}$ \\
$\Mmean$ [deg] & $124^{+44}_{-25}$ & $-89^{+72}_{-146}$ \\
\hline
\end{tabular}
\end{center}
\end{table}

The bottom row corresponds to the lower value of $\sigma = 0.001$. For this prior there exists only the anti-aligned mode (understood as a maximum of the posterior probability distribution). Predicted masses are higher with respect to the previous case, and the eccentricities are lower. For lower $\sigma = 0.00033$ (not shown) the resulting eccentricities are even lower ($e_1 \approx e_2 \approx 0.0003$) and the masses are higher ($m_1 \approx m_2 \approx 16\,\mE$). Due to very low eccentricities the apsidal lines are poorly constrained, although, similarly to the previous case, the anti-aligned system is the most likely. {The masses obtained for $\sigma = 0.033$ and $\sigma = 0.00033$ differ by a factor of $\sim 4$ and $\gtrsim 10$ for the inner and the outer planets, respectively. The differences in the eccentricities are even larger, i.e., up to two orders of magnitude.}

While different $\sigma$ gives different masses, eccentricities and apsidal lines, the best-fitting configurations in terms of the highest posterior probabilities have different values of the standard $\Chi$, for different $\sigma$. From a set of values we chose, the lowest $\Chi$ was obtained for $\sigma = 0.001$, i.e., $\Chi = 1.532$. The parameters of that configuration are listed in Tab.~\ref{tab:params_mcmc}. The quality of the fit is slightly better when compared with the parameters obtained for the periodic configuration (see Tab.~\ref{tab:params}). Nevertheless, the masses are both in agreement between the models, so do the eccentricities. The longitudes of pericentre as well as the mean anomalies differ significantly (probably due to very small eccentricities), however, the mean longitudes are in perfect agreement, as they should be.

The mass and radius of the outer planet are very similar to the values of Uranus, only the radius is $\sim 10\%$ larger. The inner planet's mass and radius suggest the composition based mainly on water \citep[e.g.,][]{Zeng2016}{, although an existence of water at such small distance from the star could be problematic.} We stress again that the masses, eccentricities as well as the relative orientation of the apsidal lines depend strongly on an a~priori information on the eccentricities. Such different orbital and physical parameters bring very different boundary conditions for the planetary systems formation theories, both in the aspect of the orbital characteristics as well as the internal structure of the planets.

\section{Conclusions}

We showed that a periodic configuration of two planets, that is a natural outcome of the migration, can be a good model of the TTV of the Kepler-25 planetary system. The period ratio of the system $P_2/P_1 = 2.039$ is significantly shifted from the nominal value of the 2:1~mean motion resonance, what may suggest a non-resonant nature of the system. We demonstrated that an anti-aligned resonant system produces the same (O-C)-diagrams as an aligned non-resonant configuration, however, the latter needs to fulfil certain criteria, like the orientation of the apsidal lines, or the relation between the eccentricities, in order to mimic the TTV signal of the periodic configuration. Due to low eccentricities of the resonant system of this value of $P_2/P_1$ ($e_1 \sim 0.0015$,  $e_2 \sim 0.0002$), as well as a degeneracy of the model mentioned above, finding a resonant configuration would be difficult (even if it was the true configuration of this system), without an a~priori information on the eccentricities. 

We studied the probability that a non-resonant configuration mimics a periodic system. Although, such a non-resonant configuration can explain the TTV produced by the periodic system, the probability is lower when the configuration is further from the periodic system. We conclude that the real architecture of the Kepler-25 planetary system is very likely resonant {in terms of librating resonant angles}\footnote{{We do not check the dynamical neighbourhood of the system, that could show whether or not the system lies in a region separated by a separatrix from the rest, non-resonant part of the phase space.}}, not only because of the probability test, but also because such a configuration is a natural outcome of the disc-induced migration, both convergent and divergent, what is believed to act a crucial role in the formation of the planetary systems.

We illustrated the dependence of the final orbital structure of the system as well as planets' masses on the assumed a~priori information on the eccentricities. Wide prior probability distributions lead to aligned orbits and small masses, while narrower priors result in anti-aligned orbits and larger masses. 
The mass-eccentricity anti-correlation shown here in the series of MCMC experiments (Fig.~\ref{fig:mcmc.new}) results from the degeneracy of the TTV signals that was illustrated with a help of an analytic model of a near-resonant two-planet system in \citep{Hadden2014}. On the other hand, the dependence of $\Delta\varpi$ (aligned/anti-aligned orbits) on the priors \citep[discussed also in][]{JontofHutter2016} can be understood from Fig.~\ref{fig:e1e2_scan}, i.e., the systems that lie along the line with $\Deltaaph = 0$ have the same TTV amplitudes and the period. Both the degeneracies make the standard MCMC modelling a challenging task and additional knowledge on the formation mechanisms (and their expected outcomes) may be very useful when setting the eccentricity priors, that are crucial for determining the systems' parameters.

We argue that aligned configurations with relatively high eccentricities that seem to be common among analysed \kepler{} systems may be artefacts and we believe that other systems with clear TTV signals should be verified in terms of their closeness to the periodic configurations, as true nature of the systems is essential for our understanding of the planetary systems formation, both their orbital configurations as well as physical compositions of the planets.

\section*{acknowledgements}

We would like to thank the anonymous reviewer for comments and suggestions that helped to improve the manuscript.
This work was partially supported by Polish National Science Centre MAESTRO grant DEC-2012/06/A/ST9/00276. 
K.G. thanks the staff of the Pozna\'n Supercomputer and Network Centre (PCSS, Poland) for
the generous and continuous support, and for providing computing
resources (grant No. 313).

\bibliographystyle{mn2e}
\bibliography{ms}
\label{lastpage}
\end{document}